\documentclass[acmsmall,screen, nonacm]{acmart}

\usepackage{amsmath}
\usepackage[T1]{fontenc}
\usepackage{pifont}
\usepackage{array}
\usepackage{bbding}
\usepackage{listings}
\usepackage{epic,eepic}
\usepackage{url}
\usepackage{float}
\usepackage{caption}
\usepackage{algpseudocode}
\usepackage{wrapfig}
\usepackage{adjustbox}
\usepackage{lipsum}
\usepackage[linesnumbered,ruled]{algorithm2e}
\usepackage{outlines}
\usepackage{color}
\usepackage{colortbl}
\usepackage{makecell} 
\usepackage{multirow}
\usepackage{xhfill}
\usepackage{textcomp}
\usepackage{booktabs}   
\usepackage{subcaption} 
\usepackage[percent]{overpic}
\usepackage{xspace}
\usepackage{empheq}
\usepackage{listings}
\usepackage{centernot}
\usepackage{xcolor}

\usepackage{amssymb}

\usepackage{graphicx}
\usepackage[inline]{enumitem}
\usepackage{tikz}
\usetikzlibrary{arrows.meta,positioning,calc,decorations.pathreplacing,patterns,fit,shapes.geometric}
\usepackage{afterpage}
\usepackage[htt]{hyphenat}
\usepackage{tcolorbox}
\usepackage{letltxmacro}
\usepackage{etoolbox}

\DeclareMathAlphabet{\cmcal}{OMS}{cmsy}{m}{n}

\usepackage{pifont}
\usepackage{xcolor}
\newcommand{\cmark}{\textcolor{green!60!black}{\ding{51}}}
\newcommand{\xmark}{\textcolor{red!75!black}{\ding{55}}}

\newcommand{\ie}{i.e.\xspace}

\newcommand{\eg}{e.g.\xspace}

\definecolor{cgreen}{RGB}{28,130,55}
\definecolor{cred}{RGB}{190,30,45}
\definecolor{corange}{RGB}{205,105,20}
\definecolor{cblue}{RGB}{40,95,175}
\definecolor{lightred}{RGB}{252,232,232}
\definecolor{lightgreen}{RGB}{232,248,235}
\definecolor{lightorange}{RGB}{254,242,226}
\definecolor{lightblue}{RGB}{218,232,252}

\newcommand{\Para}[1]{\par\addvspace{2pt}\noindent\textbf{#1}\ }

\def\compactify{\itemsep=0pt \topsep=0pt \partopsep=0pt \parsep=0pt \leftmargin=12pt}
\let\latexusecounter=\usecounter

\newcommand{\Sec}[1]{\S\ref{sec:#1}\xspace}

\usepackage{listings}
\lstdefinestyle{bpfstyle}{
  basicstyle=\ttfamily\small,
  numbers=left,
  numberstyle=\tiny\color{gray},
  numbersep=8pt,
  frame=none,
  columns=fullflexible,
  keepspaces=true,
  commentstyle=\color{gray}\itshape,
  escapeinside={(*@}{@*)}
}

\newcommand{\nop}[1]{}

\newcolumntype{P}[1]{>{\centering\arraybackslash}p{#1}} 
\newcolumntype{M}[1]{>{\centering\arraybackslash}m{#1}} 

\newcommand{\code}[1]{\texttt{\small{{#1}}}}

\newcommand{\sentinel}{element\xspace}

\begin{document}

\title{Detecting Empty Abstractions and Pruning Dead Paths in the
  Linux eBPF Verifier with Abstract Step Functions}

\author{Harishankar Vishwanathan}
\affiliation{%
  \institution{Rutgers University}
  \country{USA}}
\email{harishankar.vishwanathan@rutgers.edu}
\author{Srinivas Narayana}
\affiliation{%
  \institution{IIT Madras and Rutgers University}
  \country{USA}}
\email{srinivas.narayana@rutgers.edu}
\author{Santosh Nagarakatte}
\affiliation{%
  \institution{Rutgers University}
  \country{USA}}
\email{santosh.nagarakatte@cs.rutgers.edu}

\begin{abstract}
The Linux kernel's eBPF verifier uses abstract interpretation to
ensure the safety of user-provided programs before execution. It
tracks each register with multiple interval domains and a bitwise
domain (tristate numbers). Refining the abstract state at conditional
branches can produce mutually contradictory constraints across
domains, indicating an unreachable path. However, the verifier lacks
an empty-set ($\bot$) representation and cannot detect when these
domains have an empty intersection, causing it to analyze dead paths.
Passing such contradictory states into domain reduction or downstream
operators violates their implicit non-emptiness precondition,
triggering domain invariant violations. This has been a multi-year
issue in the Linux kernel causing production outages, false
rejections, and state explosion.

This paper presents a sound and complete constant-time emptiness check
across the eBPF verifier's abstract domains to prune dead branches. We
introduce \emph{abstract step functions}, a general framework that
tests whether intervals and non-convex domains intersect in $O(1)$
time without scanning individual values. We develop
\texttt{tnum\_step}, a formally verified constant-time operator that
jumps to the next integer satisfying a tristate bitmask. Using
\texttt{tnum\_step}, we design a multi-domain emptiness check that
detects contradictory states during branch refinement and immediately
prunes unreachable branches, eliminating the root cause of dead-branch
invariant violations. Our technique prevents invariant violations,
state explosion, and false rejections while incurring negligible
($1.48\%$) overhead across Linux kernel selftests and production
suites such as Cilium. Our \texttt{tnum\_step} operator has been
upstreamed into the mainline Linux kernel.
\end{abstract}

\maketitle


\section{Introduction}
\label{sec:introduction}

The eBPF~\cite{lwn-ebpf} framework has transformed the Linux operating
system kernel into a programmable platform that enables user-supplied
programs to customize networking, observability, and
security~\cite{cilium_repo}. Today, eBPF is widely deployed in large
production-grade compute clusters. A key component of the eBPF
ecosystem is the eBPF {\em verifier}, an in-kernel static analyzer
that checks program safety before incorporation into the kernel
address space. The verifier enforces memory safety, termination, and
type safety through abstract interpretation~\cite{cousot-1977}.

This paper concerns the static analysis in the eBPF verifier
(\Sec{background}). Along each code path in a user-supplied program,
the verifier summarizes possible register values using five
complementary \emph{abstract domains}~\cite{cousot-1977,
  cousot-1979-systematic}, which include four intervals (tracking
$[\mathit{min}, \mathit{max}]$ bounds across 64-bit and 32-bit signed
and unsigned interpretations) and a bitwise domain (tristate numbers
or tnums~\cite{regehr2006deriving, tnums-cgo22,
  fast-three-valued-arithmetic}). The tnum tracks each bit of the
64-bit register as known~\code{0}, known~\code{1}, or
unknown~(\code{x}) (\eg, $\code{01x0}_2$ represents the discrete set
$\{\code{0100}_2, \code{0110}_2\} = \{4, 6\}$).  On each instruction,
\emph{abstract operators} update these abstract values~(\eg, analyzing
\code{r1 += 10} updates $r_1$'s interval from $[2, 6]$ to $[12, 16]$).
To improve precision, the verifier propagates information between
domains through a process known as
\emph{reduction}~\cite{cousot-1979-systematic, agni-cav23} (\eg, using
the tnum $\code{01x0}_2$ allows the verifier to tighten $r_1$'s
interval bounds from $[2, 6]$ to $[4, 6]$). Together, the intervals
and tnum abstract values represent the \emph{abstract state} for a
register. Further, the verifier's abstract interpretation is {\em
  path-sensitive,} in that the verifier analyzes each code path
separately whenever it cannot statically determine that a code path
will never be taken at run time. The verifier also evolves each
register's abstract state based on the branch conditions it
encounters. For example, upon taking the branch \code{if r1 > 10 goto
  target}, the verifier refines $r_1 = [5,15]$ to $[11,15]$. Finally,
the verifier accepts a program only if every explored path passes its
safety checks.

The eBPF verifier is a key trusted component within the kernel which
enables untrusted user code to program the kernel safely yet
expressively. Hence, the verifier's soundness~\cite{bhat-2022,
  tnums-cgo22, agni-cav23, sun2024validating, sun2024finding,
  lyu2025veritas, hung2024brf, google-buzzer, agni-sas24} (\ie its
ability to reject unsafe code) and its precision~\cite{prevail-pldi19,
  nelson:exobpf:lpc:2021, sun2025prove, cai2026pcb, vayu-ebpf25} (\ie
its ability to accept safe code) have been a subject of extensive
recent research. Meanwhile, a class of bugs which has received
relatively little academic attention has remained in the kernel for
years.

\Para{Domain invariant violations in the Linux eBPF verifier.}
The eBPF verifier requires some abstract-domain-specific conditions to
hold over its abstract states, to ensure they are well-formed and
operable using its abstract operators. We call these conditions {\em
  domain invariants}. An example is ensuring $\mathit{min} \le
\mathit{max}$ in an interval abstraction. Consider a code fragment
\code{if r1 <= 10 goto target} where prior instructions constrained
$r_1$ to interval $[0, 15]$ and tnum $\mathtt{1111}_2$ (exact value
$15$). Because $r_1 = 15$, $r_1 \le 10$ is infeasible. If the branch
feasibility check inspects only the interval and not the tnum (as an
older version of the verifier did, prior to a targeted patch to fix
it~\cite{chaignon2025jeqjne}), the branch taken appears feasible, and
the branch refinement operator for \code{<=} updates $r_1$'s interval
to $[0, 10]$. Individually, $[0, 10]$ and $\mathtt{1111}_2$ look
valid, but they mutually contradict each other ($[0, 10] \cap \{15\} =
\emptyset$, or $\bot$). The verifier employed a reduction operator
which subsequently tightened the interval's lower bound using the
tnum's minimum value ($15$), producing an inverted interval $[15, 10]$
(\ie $\mathit{min} > \mathit{max}$). Instead of treating $[15, 10]$ as
$\bot$ and pruning the exploration of the path, the verifier continued
exploring it, and eventually ``recovered'' after detecting the
invariant violation, by resetting the register's bounds to $\top$, (\eg
$[0, \mathtt{UINT64\_MAX}]$ for unsigned intervals). 
These invariant violations have plagued the Linux kernel for over 
three years, triggering repeated bug
reports~\cite{sun2023jset, sun2023jsle, sjtu2025bounds, righi2026u32,
  upstream2025subreg, chaignon2025jsetselftest, krish2025} and
targeted patches across kernels v5.19--v7.1~\cite{borkmann2024fake,
  chaignon2025jset, chaignon2025crosssign, lau2025samescalar,
  lau2025tnumoverlap, chaignon2025jeqjne, chaignon2026singleton,
  eduard2026u32s32}.

Why are domain invariants violated? We observe that such violations
stem from a fundamental mismatch between conditional branch refinement
and the precondition assumed by reduction in the Linux kernel. Every
abstract ALU, jump, and reduction operator in the verifier, as well as
prior formal soundness proofs of the verifier~\cite{bhat-2022,
  agni-cav23, agni-sas24} implicitly assume \emph{concrete membership}
of the input abstract value ($\neq \bot$), \ie at least one concrete
value simultaneously satisfies all the abstract domains tracked for a
register. This condition indeed holds on paths feasible at run
time. On an infeasible branch, however, refining a register state
under a runtime-impossible condition can produce an abstract state
where each component abstract value remains syntactically well-formed
and non-empty (\eg $\mathit{min} \le \mathit{max}$), but the
concretization of the product abstract domain, \ie the intersection of
the concretizations of each component abstraction, is empty.

Even if domain invariant violations do not affect programs accepted by
the verifier at run time, they have other important consequences. A
real production outage incident reported that the Cilium container
policy-enforcement engine rejected valid networking programs upon node
reboot, bringing down connectivity for an entire Kubernetes
cluster~\cite{cilium44216, chaignon2026singleton}. Exploring paths
infeasible at run time can result in spurious program rejections,
since resetting bounds can cause the verifier to analyze unreachable
code and reject it as unsafe.  Finally, failing to prune infeasible paths can
result in {\em path explosion}, with redundant exploration of dead
code paths exhausting the verifier's budget of the total checked
instructions~\cite{lwn-million, chaignon2025jeqjne}. Our evaluations
show that a single domain invariant violation can result in an
unbounded number of programs experiencing spurious rejections or
exhausting the verifier's instruction budget.

\Para{Why is fixing domain reduction hard?}  In multi-abstract-domain
analyzers such as Frama-C/Eva~\cite{framac-2012} and
Astr\'ee~\cite{astree-2005}, computing the exact canonical reduced
product ($\rho^*$)~\cite{cousot-1979-systematic} across $k$
heterogeneous domains such as intervals, congruences, bitmasks, and
modular integer views is intractable, because materializing their
concrete sets scales exponentially in bitwidth. Further, their
geometric representations are fundamentally misaligned (convex ranges
versus non-convex sets with interior ``holes''). Consequently, practical analyzers approximate $\rho^*$ via Granger-style
pairwise reduction~\cite{granger-1992} or communication channels that
exchange coarse summary hints~(\eg, outer $\mathit{min}/\mathit{max}$
bounds)~\cite{survey-product-operators-2013}. However, these
approximations cannot detect when an interval falls inside an interior
hole of a non-convex domain, or when $k$ non-convex domains intersect
pairwise yet are globally disjoint. We are unaware of an
efficient canonical reduction operator for the combination of abstract
domains employed by the eBPF verifier.

\Para{Turning infeasibility into a branch pruning signal.}  The eBPF
verifier employs pairwise Granger reduction~\cite{granger-1992} but lacks an
explicit $\bot$ \sentinel.  We observe that empty intersections and
ill-formed states are representations of $\bot$ that provide a proof
of dead code. Because the branch refinement operators are sound on
states with concrete membership~\cite{agni-cav23, agni-sas24}, a
refined register state can become ill-formed or have an empty
intersection only if no concrete runtime value satisfies the branch
condition. By soundly (never reporting $\bot$ when a common value
exists) and completely (detecting every $\bot$ state) checking whether
each refined branch operand is ill-formed or has an empty intersection
prior to domain reduction, the analyzer (i)~ensures domain reduction
and downstream operators are never invoked on contradictory operand
states, preventing dead-branch invariant violations, and (ii)~prunes
infeasible branches before exploring them, avoiding path explosion
and spurious rejections.

\Para{Abstract step functions and set intersections.}
Our idea is to go beyond pairwise reduction, which is employed in
industrial analyzers such as Astr\'ee~\cite{astree-2005} and
Frama-C~\cite{framac-2012}.
Instead of designing a canonical reduction operator, which is
challenging for non-convex domains such as tnums (\Sec{verifier}), or
enumerating elements, our key idea is to introduce a notion of
\emph{abstract step functions}: given an
abstract value $a$ and a concrete point $z$, $\mathrm{step}(a, z)$
returns the smallest concrete member of $a$ strictly greater than $z$,
jumping across arbitrary interior gaps in a single step.
With $\mathrm{step}$, testing whether interval $[L, U]$ intersects any
non-convex domain $a$ takes at most one membership check ($L \in a$)
and one jump ($\mathrm{step}(a, L) \le U$); \eg, for $[5, 11]$ and
$\{4, 12\}$, $5 \notin \{4, 12\}$ and $\mathrm{step}(\{4, 12\}, 5) =
12 > 11$, immediately proving $\bot$.
We establish sufficient conditions for $O(1)$ $k$-domain intersection.
We develop an $O(1)$ step operator for the tnum domain,
\code{tnum\_step()}, which is formally verified with Lean and the Z3
SMT solver and upstreamed into the mainline Linux kernel.  We design a
sound and complete multi-domain $O(1)$ emptiness ($\bot$) check across
all five eBPF verifier domains using \code{tnum\_step()}. Using this
algorithm, we enable dead-path pruning before path exploration in the
verifier. It incurs negligible performance overhead ($1.48\%$ on
average) and eliminates the invariant violations that have plagued
Linux for years.

\Para{Contributions.}  (1) We characterize the abstract states of the
eBPF verifier into four classes: well-formed ($\mathit{WF}$),
ill-formed ($\mathit{IF}$),
concrete-membership ($\mathit{CM}$), and empty-intersection
($\mathit{EI}$).  We show how infeasible branches produce
empty-intersection ($\bot$) states that violate the implicit
non-emptiness precondition of domain reduction and downstream
operators, causing domain invariant violations.  We show that reaching
$\mathit{EI}$ or $\mathit{IF}$ states after branch refinement is a
sound guarantee of runtime branch infeasibility~(\S
\ref{sec:verifier}).
(2) We propose to go beyond pairwise domain reduction and formulate
abstract step functions. We design \code{tnum\_step()}, a formally
verified and $O(1)$ bitwise step operator for the tristate domain.
The algorithm has been upstreamed into the mainline Linux
kernel~(\S\ref{sec:step}).
(3) We develop a general algorithm using step functions for deciding
joint concrete membership across $k$ abstract
domains~(\S\ref{sec:general}). We design a sound and complete $O(1)$
emptiness ($\bot$) check across all five abstract domains in the eBPF
verifier~(\S\ref{sec:ebpf}).
(4) We integrate our $\bot$ check into the Linux eBPF verifier to
prune dead paths before path exploration~(\S\ref{sec:ebpf}) and
evaluate it across historical bugs, microbenchmarks, Linux kernel
selftests, and Cilium~(\S\ref{sec:eval}).

  \section{Background on the eBPF Verifier's Abstract Domains}
\label{sec:background}
We provide background on the eBPF verifier's abstract domains and the
domain reduction operator.

\Para{Concrete semantics.} Let $\mathbb{W}_w \triangleq \{0, 1,
\dots, 2^w - 1\}$ be the set of $w$-bit unsigned integers, and
$\mathbb{S}_w \triangleq \{-2^{w-1}, \dots, 2^{w-1} - 1\}$ the set of
$w$-bit two's-complement signed integers. The primary concrete
universe of register values in eBPF is the set of 64-bit machine
words, $\mathbb{W}_{64}$. For any 64-bit machine word $x \in
\mathbb{W}_{64}$, we denote the signed interpretation as
$\text{s64}(x) \in \mathbb{S}_{64}$, the lower 32-bit sub-register word
as $x_{[31:0]} \triangleq (x \ \& \ \text{\code{0xffffffff}}) \in
\mathbb{W}_{32}$, and the signed interpretation of the lower 32-bit
sub-register word as $\text{s32}(x_{[31:0]}) \in \mathbb{S}_{32}$. An
eBPF program executes on a register-based virtual machine with 11
64-bit registers $\mathcal{V} \triangleq \{r_0, \dots, r_{10}\}$.

\Para{Instructions.} We focus on arithmetic and logic~(ALU) and
conditional jump instructions. ALU instructions are of the form $r_d
\code{=} \text{op}\code{(}r_d, r_s\code{)}$ or $r_d \allowbreak
\code{=} \text{op}\code{(}r_d, imm\code{)}$, where $r_d, r_s \in
\mathcal{V}$ are the destination and source registers, $imm$ is a
constant immediate operand encoded in the instruction, and $\text{op}
\in \{\texttt{+}, \texttt{-}, \texttt{\&}, \texttt{|},
\texttt{\textasciicircum}, \texttt{<{}<}, \texttt{>{}>}, \dots\}$.
Conditional jump instructions are of the form \code{if (}$r_d\text{ op
}r_s$\code{) goto target} or \code{if (}$r_d\text{ op
}\mathit{imm}$\code{) goto target}, where $\text{op} \in
\{\texttt{==}, \texttt{!=}, \texttt{<}, \texttt{<=}, \texttt{>},
\texttt{>=}, \texttt{\&}, \dots\}$ (including both unsigned and signed
comparisons). Conditional jumps compare full 64-bit values over
$\mathbb{W}_{64}$ (or the lower 32-bit sub-register values for 32-bit
jump variants). If the comparison holds at runtime, execution jumps to
$\code{target}$; otherwise, it falls through to the next sequential
instruction.

\Para{eBPF abstract domains and register states.}
The eBPF verifier uses abstract
interpretation~\cite{cousot-1977,tnums-cgo22,agni-cav23} to
over-approximate the runtime values of each register using values from
an \emph{abstract domain}. Formally, an abstract domain $\mathbb{A}
\triangleq (A, \sqsubseteq_A, \gamma_A)$ consists of a set of abstract
values $A$, a concretization function $\gamma_A : A \to
2^{\mathbb{W}_{w}}$ mapping each abstract value to the set of $w$-bit
words it represents ($w \in \{64, 32\}$), and a partial order $a_1 \sqsubseteq_A a_2 \iff
\gamma_A(a_1) \subseteq \gamma_A(a_2)$, with top element $\top_A$
($\gamma_A(\top_A) = \mathbb{W}_{w}$) and bottom element $\bot_A$
($\gamma_A(\bot_A) = \emptyset$).

\Para{Multi-domain value tracking.}
The verifier tracks each register $r \in \mathcal{V}$ using five
abstract domains~\cite{agni-cav23}
as 2-tuples of words (denoted $[\cdot, \cdot]$ for
intervals, $(\cdot, \cdot)$ for tnums).

\begin{itemize}[noitemsep, nosep, leftmargin=*]
\item \emph{64-bit interval domains ($\mathbb{A}_{\text{u64}},
\mathbb{A}_{\text{s64}}$)}.  The unsigned domain
$\mathbb{A}_{\text{u64}}$ is a pair $[u64_{\min}, u64_{\max}] \in
\mathbb{W}_{64} \times \mathbb{W}_{64}$ with $u64_{\min} \le
u64_{\max}$ and the concretization function:
$\gamma_{\text{u64}}([u64_{\min}, u64_{\max}]) \triangleq \{ x \in
\mathbb{W}_{64} \mid u64_{\min} \le x \le u64_{\max} \}$. The signed
64-bit interval domain $\mathbb{A}_{\text{s64}}$ is a pair
$[s64_{\min}, s64_{\max}] \in \mathbb{S}_{64} \times \mathbb{S}_{64}$
with $s64_{\min} \le s64_{\max}$ and the concretization function:
$\gamma_{\text{s64}}([s64_{\min}, s64_{\max}]) \triangleq \{ x \in
\mathbb{W}_{64} \mid s64_{\min} \le \text{s64}(x) \le s64_{\max} \}$.
\item \emph{32-bit sub-register interval domains
($\mathbb{A}_{\text{u32}}, \mathbb{A}_{\text{s32}}$)}. These domains
track the unsigned and signed bounds of the lower 32 bits $x_{[31:0]}$
of a register.  An abstract value in $\mathbb{A}_{\text{u32}}$ is a
pair $[u32_{\min}, u32_{\max}] \in \mathbb{W}_{32} \times
\mathbb{W}_{32}$ with $u32_{\min} \le u32_{\max}$ and concretization
function: $\gamma_{\text{u32}}([u32_{\min}, u32_{\max}]) \triangleq \{
x \in \mathbb{W}_{64} \mid u32_{\min} \le x_{[31:0]} \le u32_{\max}
\}$.
Similarly, an abstract value in $\mathbb{A}_{\text{s32}}$ is a pair
$[s32_{\min}, s32_{\max}] \in \mathbb{S}_{32} \times \mathbb{S}_{32}$
with $s32_{\min} \le s32_{\max}$ and concretization function:
$\gamma_{\text{s32}}([s32_{\min}, s32_{\max}]) \triangleq \{ x \in
\mathbb{W}_{64} \mid s32_{\min} \le \text{s32}(x_{[31:0]}) \le
s32_{\max} \}$.
\item \emph{Tristate domain ($\mathbb{A}_{\mathit{tnum}}$)}. Tracks
known ($0$/$1$) and unknown bits across all 64 bit positions of a
register. An abstract value is a pair $(v, m) \in \mathbb{W}_{64}
\times \mathbb{W}_{64}$, with $v \mathbin{\&} m = 0$, where $v$
(\code{value}) denotes known bits, $m$ (\code{mask}) denotes
unknown bits, and `$\&$' denotes bitwise logical
`and'. The concretization is $\gamma_{\mathit{tnum}}((v, m))
\triangleq \{x \in \mathbb{W}_{64} \mid (x \mathbin{\&} \mathord{\sim}
m) = v\}$. 
\end{itemize}
An {\em abstract operator} $g: \mathbb{A} \times \mathbb{A} \to
\mathbb{A}$ abstracts a concrete operation $f: \mathbb{W} \times
\mathbb{W} \to \mathbb{W}$, permitting efficient program analysis
directly over abstract values $a_1, a_2 \in \mathbb{A}$. The abstract
operator is {\em sound} if $f(\gamma(a_1), \gamma(a_2)) \subseteq
\gamma(g(a_1, a_2))$~\cite{mine-tutorial-2017}.

\Para{Domain invariants.}  We term the conditions
under which an abstract value in a domain is
well-defined (\eg $s32_{min} \le s32_{max}$ for
$\mathbb{A}_{s32}$ and $v\ \&\ m = 0$ for
$\mathbb{A}_{\mathit{tnum}}$) {\it domain
  invariants.}
Each domain $\mathbb{A}$ in the verifier's kernel
implementation (\code{struct
bpf\_reg\_state}) lacks an explicit
$\bot_{\mathbb{A}}$ representation. 
Every well-defined abstract value represents a
non-empty set, \ie
$\gamma_{\mathbb{A}}(a_{\mathbb{A}}) \neq \emptyset$.

\Para{Product domain and domain reduction.}
The verifier combines these five domains into the product domain
$\mathbb{P} \triangleq \mathbb{A}_{u64} \times \mathbb{A}_{s64} \times
\mathbb{A}_{u32} \times \mathbb{A}_{s32} \times
\mathbb{A}_{\mathit{tnum}}$. An abstract state $R \triangleq
(a_{\text{u64}}, a_{\text{s64}}, a_{\text{u32}}, a_{\text{s32}},
a_{\mathit{tnum}}) \in \mathbb{P}$ has a component-wise partial order
$\sqsubseteq_{\mathbb{P}}$. The concretization of $R$ is the set of
64-bit words that simultaneously satisfy all five component domains:
$\gamma_{\mathbb{P}}(R) \triangleq \gamma_{\text{u64}}(a_{\text{u64}})
\cap \gamma_{\text{s64}}(a_{\text{s64}}) \cap
\gamma_{\text{u32}}(a_{\text{u32}}) \cap
\gamma_{\text{s32}}(a_{\text{s32}}) \cap
\gamma_{\mathit{tnum}}(a_{\mathit{tnum}})$. \emph{Domain
reduction}~\cite{cousot-1979-systematic,
  survey-product-operators-2013} applies an operator $\rho :
\mathbb{P} \to \mathbb{P}$ that refines the individual components of
$R$ (\ie, $\rho(R)_i \sqsubseteq_{i} R_i$ for some component
$i$) without changing the underlying concrete set
($\gamma_{\mathbb{P}}(\rho(R)) = \gamma_{\mathbb{P}}(R)$).

Ideally,
reduction computes the canonical \emph{reduced
product}~\cite{cousot-1979-systematic} $\rho^*(R) \triangleq
\bigsqcap_{\mathbb{P}}\{R' \in \mathbb{P} \mid \gamma_{\mathbb{P}}(R)
\subseteq \gamma_{\mathbb{P}}(R')\}$, which is the most precise
abstract state in $\mathbb{P}$ representing $\gamma_{\mathbb{P}}(R)$.
Since computing $\rho^*$ directly is typically
expensive~\cite{survey-product-operators-2013}, the verifier follows
Granger's approach~\cite{granger-1992, agni-cav23}, composing pairwise reduction operators
$\rho_{i \to j}$ between domains. For instance, $\rho_{\mathit{tnum}
\to u64}$ clamps the unsigned interval to $[\min(a_{\mathit{tnum}}),$
$\max(a_{\mathit{tnum}})]$ and $\rho_{u64 \to \mathit{tnum}}$ updates
the tnum with the common high-order bit prefix of $u64_{\min}$ and
$u64_{\max}$. Because these pairwise operators exchange only outer
bounds and prefixes, reduction is incomplete: it can fail to reach
$\rho^*(R)$ or detect when $\gamma_{\mathbb{P}}(R) = \emptyset$.
For example, consider the two-domain state $R =
(a_{\text{u64}}, a_{\mathit{tnum}}) = ([10, 14], \mathtt{1x0x}_2)$.
Here, $[10, 14]$ admits $\{10, \ldots, 14\}$ and $\mathtt{1x0x}_2$
admits $\{8, 9, 12, 13\}$, so their intersection is
$\gamma_{\mathbb{P}}(R) = \{12, 13\}$ and the canonical reduced
product is $\rho^*(R) = ([12, 13], \mathtt{110x}_2)$. Under pairwise
reduction, however, $\rho_{\mathit{tnum} \to \text{u64}}$ clamps $[10,
14]$ to the tnum's outer span $[8, 13]$ to obtain $[10, 13]$, and
$\rho_{\text{u64} \to \mathit{tnum}}$ observes that $10 =
\mathtt{1010}_2$ and $13 = \mathtt{1101}_2$ share only the most
significant bit ($\mathtt{1xxx}_2$), leaving $\mathtt{1x0x}_2$
unchanged. Reduction thus stops at $([10, 13], \mathtt{1x0x}_2)$,
which is strictly less precise than $\rho^*(R) = ([12, 13],
\mathtt{110x}_2)$.

\Para{Abstract operators and reductions in the
  Linux eBPF verifier.}  The Linux kernel (as of
version 7.1-rc1) implements abstractions and
reductions in a few key functions.
On an ALU
instruction $r_d \leftarrow \mathrm{op}(r_d, r_s)$, the verifier
applies the abstract operator for $\mathrm{op}$
(\code{adjust\_reg\_min\_max\_vals()})
to over-approximate
$\mathrm{op}$ in $r_d$'s abstract state $R_d$, followed by reduction
(\code{reg\_bounds\_sync()}).
On a conditional jump \code{if (}$r_1\text{ op }r_2$\code{) goto
target} with condition $C$, the verifier performs two steps: (1)~it
checks whether $C$ is statically always true, always false, or unknown,
(\code{is\_branch\_taken()}),
exploring only the taken (true) or
fall-through (false) path when $C$ is known, and both paths when $C$
is unknown; (2)~along each explored path, it refines the operand
register states $(R_1, R_2)$ to $(R_{1t}, R_{2t})$ under $C$ or
$(R_{1f}, R_{2f})$ under $\neg C$,
(\code{regs\_refine\_cond\_op()}),
followed by reduction.

  \section{Domain Invariant Violations and the Dead Branch Principle}
\label{sec:verifier}

The abstract operators in Linux eBPF have been checked in prior formal
verification efforts~\cite{agni-cav23, lyu2025veritas, bhat-2022,
  shung-hsi-model-checking-tnums} as well as through systematic
fuzzing to detect mismatches between concrete executions and
abstractions~\cite{sun2024validating, aee-atc25, hung2024brf,
  sun2024finding}. Despite extensive scrutiny, the eBPF verifier still
suffers from {\em violations of domain invariants}, \eg producing
abstract values with $u64_{min} > u64_{max}$ in $\mathbb{A}_{\textit{u64}}$
(\Sec{background}).

\Para{Why did prior soundness verification miss domain invariant violations?}
We consider Agni~\cite{agni-cav23}, a formal
verification framework which proved that
BPF's ALU and jump operators are sound (verified up to
Linux v7.1-rc1). Agni formalized conditional jump soundness as shown
in Equation~(\ref{eqn:agni-soundness}), where $\code{refine}(R_1, R_2,
C)$ (short for \code{regs\_refine\_cond\_op()} in
the kernel (\Sec{background})) refines
the input abstract states $R_1$ and $R_2$ under condition $C$ (or
$\neg C$) to produce $(R_{1t}, R_{2t})$ (or $(R_{1f}, R_{2f})$).

{\small
\begin{equation}
\begin{aligned}
  &\forall R_1, R_2 \in \mathbb{P},\; x_1 \in \gamma_{\mathbb{P}}(R_1),\; x_2 \in \gamma_{\mathbb{P}}(R_2) :\;
  (R_{1t}, R_{2t}) = \mathtt{refine}(R_1, R_2, C) \;\land\; (R_{1f}, R_{2f}) = \mathtt{refine}(R_1, R_2, \neg C) \\[-2pt]
  &\quad \implies \big(C(x_1, x_2) \implies x_1 \in \gamma_{\mathbb{P}}(R_{1t}) \land x_2 \in \gamma_{\mathbb{P}}(R_{2t})\big)
   \;\land\; \big(\neg C(x_1, x_2) \implies x_1 \in \gamma_{\mathbb{P}}(R_{1f}) \land x_2 \in \gamma_{\mathbb{P}}(R_{2f})\big)
\end{aligned}
\label{eqn:agni-soundness}
\end{equation}}%

Equation~(\ref{eqn:agni-soundness}) guarantees that live
branches preserve all runtime values, but does not enforce domain
invariants on {\em runtime-infeasible paths}. Consider the case where $C(x_1, x_2)$
is false for all $x_1 \in \gamma_{\mathbb{P}}(R_1), x_2 \in
\gamma_{\mathbb{P}}(R_2)$ (\ie, the true path is dead). When the
verifier fails to statically determine that $C(x_1, x_2)$ is always
false, it explores the true path and invokes $\mathtt{refine}(R_1,
R_2, C)$. Because the premise $C(x_1, x_2)$ is false for all concrete
pairs, $C(x_1, x_2) \!\!\implies\!\! x_1\! \in
\!\gamma_{\mathbb{P}}(R_{1t}) \land x_2\! \in\!
\gamma_{\mathbb{P}}(R_{2t})$ holds \emph{vacuously}: the operator
remains sound under
Equation~(\ref{eqn:agni-soundness}). Moreover, no well-formedness
constraints are placed on $R_{1t}, R_{2t}$.
The problem is not unique to Agni. A straightforward logical encoding
of abstract operator soundness (\Sec{background}) typically includes the
premise of  concrete membership in the input abstract
value~\cite{taneja2020testing, shung-hsi-model-checking-tnums,
  tnums-cgo22, bhat-2022}. Fuzzers relying primarily on run-time
checks~\cite{sun2024validating, aee-atc25, hung2024brf,
  sun2024finding} (even if specification-guided~\cite{lyu2025veritas})
will also miss domain invariant violations.

In the eBPF verifier, refining an abstract value under an impossible
condition $C$ may produce a
``contradictory'' register state $R$ with $\gamma_{\mathbb{P}}(R) =
\emptyset$. Because the verifier lacks an explicit $\bot$ \sentinel, domain
reduction and downstream operators assume concrete membership
($\gamma_{\mathbb{P}}(R) \neq \emptyset$) as a precondition. When
invoked on a contradictory state, these operators
can compute ill-formed abstract values (\eg, $u64_{\min} > u64_{\max}$), triggering
domain invariant violations.
Adding an explicit $\bot$ \sentinel is easy, but deciding when an
abstract operator should emit $\bot$ is not as straightforward.

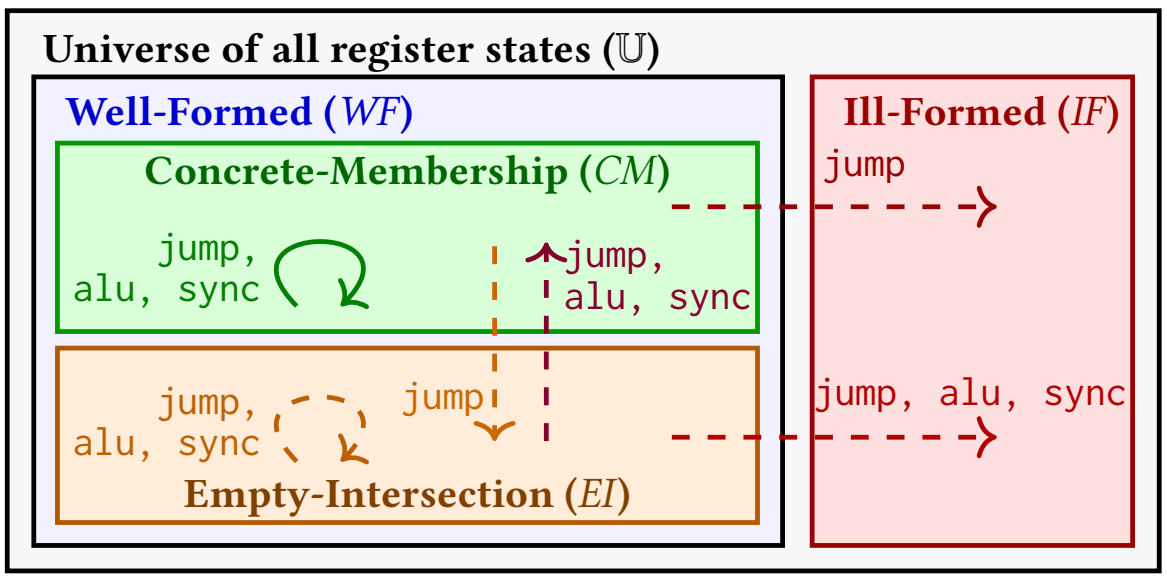
\begin{wrapfigure}{r}{0.55\textwidth}
\centering
\resizebox{\linewidth}{!}{%
\begin{tikzpicture}[x=0.87cm, y=0.75cm]

    \draw[thick, fill=black!3] (0,0) rectangle (7.80, 4.4);
    \node[anchor=north west, font=\scriptsize\bfseries] at (0.1, 4.4)
        {Universe of all register states ($\mathbb{U}$)};

    \draw[thick, fill=blue!5] (0.18, 0.2) rectangle (5.25, 3.88);
    \node[anchor=north west, font=\scriptsize\bfseries, text=blue!80!black] at
        (0.25, 3.9) {Well-Formed ($\mathit{WF}$)};

    \draw[thick, fill=green!15, draw=green!60!black] (0.34, 1.88) rectangle (5.08, 3.36);
    \node[anchor=center, font=\scriptsize\bfseries, text=green!40!black]
        at (2.71, 3.12) {Concrete-Membership ($\mathit{CM}$)};

    \draw[thick, fill=orange!15, draw=orange!70!black] (0.34, 0.38) rectangle (5.08, 1.75);
    \node[anchor=center, font=\scriptsize\bfseries, text=orange!50!black]
        at (2.71, 0.58) {Empty-Intersection ($\mathit{EI}$)};

    \draw[thick, fill=red!12, draw=red!60!black] (5.45, 0.2) rectangle (7.62, 3.88);
    \node[anchor=north west, font=\scriptsize\bfseries, text=red!60!black]
        at (5.52, 3.9) {Ill-Formed ($\mathit{IF}$)}; 

    \draw[->, thick, green!50!black] (1.96, 2.08) to[out=135, in=45, looseness=7] (2.26, 2.08);
    \node[anchor=east, font=\scriptsize, text=green!50!black, align=right] at (1.86, 2.36) {
        \texttt{jump,}\\[-1pt]
        \texttt{alu, sync}
    };

    \draw[->, dashed, thick, orange!75!black] (1.96, 0.86) to[out=135, in=45, looseness=7] (2.26, 0.86);
    \node[anchor=east, font=\scriptsize, text=orange!75!black, align=right] at (1.86, 1.15) {
        \texttt{jump,}\\[-1pt]
        \texttt{alu, sync}
    };

    \draw[->, dashed, thick, orange!80!black] (3.3, 2.55) -- (3.3, 1.02);
    \node[anchor=east, inner xsep=1.5pt, font=\scriptsize, text=orange!80!black] at (3.3, 1.35) {\texttt{jump}};

    \draw[->, dashed, thick, purple!70!black] (3.65, 1.02) -- (3.65, 2.58);
    \node[anchor=west, inner xsep=1.5pt, font=\scriptsize, text=purple!70!black, align=left] at (3.7, 2.30) {
        \texttt{jump,}\\[-1pt]
        \texttt{alu, sync}
    };

    \draw[->, dashed, thick, red!60!black] (4.5, 2.86) -- (6.7, 2.86);
    \node[anchor=south, font=\scriptsize, text=red!60!black] at (5.8, 2.88) {\texttt{jump}};

    \draw[->, dashed, thick, red!70!black] (4.5, 1.05) -- (6.7, 1.05);
    \node[anchor=south, font=\scriptsize, text=red!70!black] at (6.52, 1.08) {\texttt{jump, alu, sync}};

\end{tikzpicture}%
}
\caption{\textbf{Abstract State Partition and Transition Dynamics.} 
Partition of $\mathbb{U}$ into $\mathit{CM}, \mathit{EI},
\mathit{IF}$. Arrows show transitions grouped by operator category
(solid: runtime-feasible paths, dashed: runtime-infeasible; SMT-verified).}
\label{fig:operator_transitions}
\end{wrapfigure}

\Para{Characterizing the verifier's abstract state space.}
We define the \emph{syntactic universe} of all 10-tuples of
machine words in \code{struct bpf\_reg\_state} as $\mathbb{U}
\triangleq \mathbb{W}_{64}^2 \times \mathbb{S}_{64}^2 \times
\mathbb{W}_{32}^2 \times \mathbb{S}_{32}^2 \times \mathbb{W}_{64}^2$.
Not every representable tuple $R \in \mathbb{U}$ corresponds to a
well-formed abstract state. Further, the eBPF verifier does not define a
$\bot$ \sentinel to represent the empty set. To clearly identify the
states that arise with infeasible paths, we partition $\mathbb{U}$
into three classes (Figure~\ref{fig:operator_transitions}).

\textbf{(1) Well-formed states ($\mathit{WF}$).} An abstract register
state $R \in \mathbb{U}$ is \emph{well-formed} ($R \in \mathit{WF}$)
if and only if each component domain $d \in \mathcal{D} \triangleq
\{\text{u64}, \text{s64}, \text{u32}, \text{s32}, \text{tnum}\}$
satisfies its domain-specific invariant
(\eg $v\ \&\ m = 0$ for
$\mathbb{A}_{\mathit{tnum}}$, see \Sec{background}), and thus independently
represents a non-empty set of concrete words: $R \in \mathit{WF} \iff
\bigwedge_{\mathbb{A} \in \mathcal{D}} \gamma_{\mathbb{A}}(a_{\mathbb{A}}) \neq \emptyset$.

The $\mathit{WF}$ set is further partitioned into two mutually
disjoint, exhaustive subsets:

\textbf{(1)(a) Concrete-membership states ($\mathit{CM}$).} A
well-formed register state $R \in \mathit{WF}$ is a
concrete-membership state ($R \in \mathit{CM}$) if and only if
at least one concrete 64-bit word $x \in \mathbb{W}_{64}$
simultaneously belongs to all five component domains: $R \in
\mathit{CM} \iff R \in \mathit{WF} \;\land\; \gamma_{\mathbb{P}}(R)
\neq \emptyset$.

\textbf{(1)(b) Empty-intersection states ($\mathit{EI}$).} A well-formed
register state $R \in \mathit{WF}$ is an empty-intersection
state ($R \in \mathit{EI}$) if each component domain is individually
well-formed, but conveys contradictory information as the mutual
intersection is empty: $R \in
\mathit{EI} \iff R \in \mathit{WF} \;\land\; \gamma_{\mathbb{P}}(R) =
\emptyset$.
Whereas canonical reduction $\rho^*$ would reduce $\mathit{EI}$ states
to $\bot$, the eBPF verifier's pairwise domain
reduction (\code{reg\_bounds\_sync()}) only
exchanges outer bounding boxes and
shared bit prefixes, leaving $\mathit{EI}$ states
unreduced (\Sec{background} shows an example).

\textbf{(2) Ill-formed states ($\mathit{IF}$).} A register state $R \in
\mathbb{U}$ is ill-formed ($R \in \mathit{IF}$) if at least one
component domain violates its domain invariant: $R
\in \mathit{IF} \iff  \bigvee_{\mathbb{A} \in \mathcal{D}} \gamma_{\mathbb{A}}(a_{\mathbb{A}}) =
\emptyset$.
For any $R \in \mathit{IF}$, the component domain $\mathbb{A}$ violating its
well-formedness invariant represents a component-level bottom element
$\bot_{\mathbb{A}}$. However, because the eBPF verifier's domains lack a valid
representation of $\bot_{\mathbb{A}}$, when the well-formedness checker
(\code{reg\_bounds\_sanity\_check()}) detects $R \in \mathit{IF}$, it
emits a domain invariant violation diagnostic message and resets each
component domain to $\top_{\mathbb{A}}$.

Together, $\mathit{EI} \cup \mathit{IF}$ comprises all states in
$\mathbb{U}$ whose joint concretization is empty (\ie,
$\gamma_{\mathbb{P}}(R)  = \emptyset$). Further, the sets
$\mathit{CM}$, $\mathit{EI}$, and $\mathit{IF}$ form an exact,
mutually disjoint partition of the entire universe: $\mathbb{U} =
\mathit{CM} \uplus \mathit{EI} \uplus \mathit{IF}$. Every register
state tuple $R \in \mathbb{U}$ belongs to exactly one of
$\mathit{CM}$, $\mathit{EI}$, or $\mathit{IF}$.

\Para{State transitions.} Figure~\ref{fig:operator_transitions}
illustrates how abstract operators evolve register states across
$\mathit{CM}$, $\mathit{EI}$, and $\mathit{IF}$ (solid: runtime-feasible
paths; dashed: runtime-infeasible), revealing five key behaviors that
we have formally checked through an SMT encoding of the kernel's
abstract operators (7.1-rc1).

\begin{enumerate}[leftmargin=*,noitemsep]
\item {\textit{Operator soundness (\eg as defined by
    Agni~\cite{agni-cav23}) holds on live paths}
($\mathit{CM} \to \mathit{CM}$)}.  Feasible jumps map $\mathit{CM}$
inputs to valid $\mathit{CM}$ outputs
(Equation~\ref{eqn:agni-soundness}); ALU ops and domain reduction
(\code{reg\_bounds\_sync()}) likewise preserve $\mathit{CM}$.

\item \textit{Only infeasible branches escape $\mathit{CM}$}
($\mathit{CM} \to \mathit{EI}$, $\mathit{CM} \to \mathit{IF}$). Along
infeasible paths, jumps can produce contradictory well-formed
states ($\mathit{EI}$) or invariant-violating abstractions ($\mathit{IF}$).

\item \textit{Hidden contradictions can propagate} ($\mathit{EI} \to
\mathit{EI}$). Domain reduction (\code{reg\_bounds\_sync()}) leaves
intervals in tnum gaps unreduced (\eg, $([5, 11], \mathtt{x100}_2)$),
and jump or ALU ops (\eg, \code{add}) can propagate $\mathit{EI}$
states.

\item \textit{Contradictory states can trigger domain invariant
  violations} ($\mathit{EI} \to
\mathit{IF}$). Jumps (\eg, \code{jeq/jne}), ALU ops (\eg, \code{or},
\code{and}, \code{lsh}, \code{arsh}), and domain reduction
(\code{reg\_bounds\_sync()}) can all transition $\mathit{EI}$ inputs
into $\mathit{IF}$.

\item \textit{Spurious recovery} ($\mathit{EI} \to \mathit{CM}$).
Jumps (\eg, \code{jset} resetting ranges~\cite{chaignon2025jset}), ALU
ops (\eg, \code{rsh}, \code{and}, \code{add}), and domain reduction
(\code{reg\_bounds\_sync()}) can transition $\mathit{EI}$ back to
$\mathit{CM}$.
\end{enumerate}

Transitions into $\mathit{IF}$ via (2) and (4) trigger domain
invariant violation diagnostics, whereas silent propagation (3) and
spurious recovery (5) let dead branches persist, which cause delayed
violations, state explosion, or false rejections on downstream dead
code (\Sec{introduction}).

In classical abstract
interpretation~\cite{cousot-1977, cousot-1979-systematic,
mine-tutorial-2017}, a conditional branch guarded by condition $C$ is
modeled by an abstract guard transfer function $\text{g}_C^\sharp:
\mathbb{A} \to \mathbb{A}$. A sound guard
operator must over-approximate all concrete values satisfying the
condition: $\{x \in \gamma(a) \mid C(x)\} \subseteq
\gamma(\text{g}_C^\sharp(a))$. Whenever $\text{g}_C^\sharp(a) = \bot$,
soundness implies that no concrete runtime value can
satisfy the condition ($\{x \in \gamma(a) \mid C(x)\} = \emptyset$),
certifying that the branch is unreachable and can be soundly pruned.
In a multi-domain product analyzer (such as
Astr{\'e}e~\cite{astree-2005} or Frama-C~\cite{framac-2012}), each
domain refines its abstract bounds and invokes domain reduction
$\rho$, pruning the path if the joint state reduces to $\bot$.
In the eBPF verifier,
the refinement operator for jumps acts as the guard. eBPF currently
lacks the $\bot$ \sentinel, and more importantly, lacks a
canonical reduction operator to conclude that an abstract state after
a conditional branch should be $\bot$.
To the best of our knowledge, we
are unaware of efficient canonical reduction operators for the
combination of eBPF's abstract domains (\Sec{background}).

\Para{Dead branch principle.} Taking an alternative approach, the
\emph{Dead Branch Principle}, formalized below, states that reaching
$\mathit{EI} \cup \mathit{IF}$ in the eBPF verifier after branch
refinement is the exact operational equivalent of reducing to
$\bot$. Since only branches escape CM, detecting abstract
states $a \in \mathit{EI} \cup \mathit{IF}$ and pruning branches that
reach such states prevents domain invariant violations.

\begin{theorem}[Dead Branch Principle]
\label{thm:dead_branch}
Let $P$ be an eBPF program, and let $e = (u, v)$ be a conditional jump
edge at $u$ with condition $C$. Let $R_u \in \mathit{CM}$ be the
abstract state of an operand register prior to the branch. If the branch
refinement operator along edge $e$ produces an abstract state $R_v \in
(\mathit{EI} \cup \mathit{IF})$, then edge $e$ is dead: no concrete
execution of $P$ reaching $u$ can take edge $e$ at runtime.
\end{theorem}

\begin{proof}
We proceed by contradiction, assuming edge $e$ is live. Then there
exists a concrete value $x \in \gamma_{\mathbb{P}}(R_u)$ satisfying
$C(x) = \mathit{true}$. By operator soundness on live paths
(Equation~\ref{eqn:agni-soundness}), any value $y \in
\gamma_{\mathbb{P}}(R_u)$ satisfying $C(y)$ must belong to the output
state: $\{y \in \gamma_{\mathbb{P}}(R_u) \mid C(y)\} \subseteq
\gamma_{\mathbb{P}}(R_v)$. Since $x$ satisfies both conditions, $x \in
\gamma_{\mathbb{P}}(R_v)$, which implies $\gamma_{\mathbb{P}}(R_v)
\neq \emptyset$. By hypothesis, however, $R_v \in (\mathit{EI} \cup
\mathit{IF})$, meaning $\gamma_{\mathbb{P}}(R_v) = \emptyset$ by
definition, a contradiction. Hence, edge $e$ must be dead.
\end{proof}

  \section{Abstract Step Functions}
\label{sec:step}

Theorem~\ref{thm:dead_branch} and \Sec{verifier} show that detecting
$R \in \mathit{EI} \cup \mathit{IF}$ suffices to prevent domain
invariant violations. While checking $R \in \mathit{IF}$ is easy,
deciding $R \in \mathit{EI}$ requires testing whether the domains
share any concrete member ($\gamma_{\mathbb{P}}(R) \neq
\emptyset$). We now motivate why convex intersection techniques fail
over bitwise domains (\S\ref{sec:need_step}), and introduce abstract
step functions which enable efficient procedures to detect empty
concrete intersections of abstract values (\Sec{tnum_step}).

\subsection{Need for Step Functions in Detecting Empty Set Intersections}
\label{sec:need_step}

We are interested in efficient decision procedures to detect the
emptiness of the intersection of concretizations of several abstract
values. We regard such a procedure {\em sound} if the concrete
intersection is indeed empty whenever the procedure returns empty, and
{\em complete} if the procedure returns empty whenever the concrete
intersection is empty.

The interval abstraction admits a sound, complete, and efficient
procedure to check intersection emptiness: $\bigcap_{i=1}^n [L_i, U_i]
\neq \emptyset \iff \max(L_1, \dots, L_n) \le \min(U_1, \dots,
U_n)$. The convexity of intervals in one-dimension, \ie the absence of
internal ``holes'' in the concretization, permits the construction of
such a procedure by relying on just the interval boundaries. However,
in the eBPF verifier, $\mathbb{P}$ is non-convex. A tristate number
with $k$ unknown bits ($k$ < bitwidth) may represent up to $2^k$
non-contiguous integers (\eg $\gamma(\mathtt{x0x1}_2) = \{1, 3, 9,
11\}$). Approximating $t = (v, m) \in \mathbb{A}_{\mathit{tnum}}$ by
its bounding box $[t_{\min}, t_{\max}] = [v, v \mathbin{|} m]$, and
running the interval emptiness check above is sound but
incomplete. For example, interval $[7, 9] = \{7, 8, 9\}$ and tnum $(4,
9) = \mathtt{x10x}_2 = \{4, 5, 12, 13\}$ have overlapping bounding
boxes $[7, 9]$ and $[4, 13]$, yet their intersection is empty because
$[7, 9]$ falls inside the internal gap $(5, 12)$ of $t$. To detect
empty intersections soundly and completely without enumerating the
concrete elements in $\mathbb{C}$, we introduce an operator that jumps
over an internal hole in the concretization. We show later how to
build on this operator to construct efficient procedures to check the
emptiness of intersections.

\begin{definition}[Abstract Step Function]
\label{def:step}
Let $\mathbb{A}$ be an abstract domain over a totally ordered set
$(\mathbb{C}, \le)$ with concretization $\gamma_{\mathbb{A}} :
\mathbb{A} \to 2^{\mathbb{C}}$. The \emph{abstract step function}
$\mathrm{step}_{\mathbb{A}} : \mathbb{A} \times \mathbb{C} \to
\mathbb{C} \cup \{\infty\}$ returns the smallest member of
$\gamma_{\mathbb{A}}(a)$ strictly greater than $x$:%
\[
\mathrm{step}_{\mathbb{A}}(a, x) \triangleq \min\{y \in \gamma_{\mathbb{A}}(a) \mid y > x\} \quad (\text{with } \min \emptyset \triangleq \infty).
\]
\end{definition}
A corollary is that if $w \in \gamma_{\mathbb{A}}(a)$ and $w > x$,
then $x < \mathrm{step}_{\mathbb{A}}(a, x) \leq w$.
For contiguous intervals $[L, U]$, the step function is simple:
$\mathrm{step}([L, U], x) = \max(L, x + 1)$ if $x < U$ and $\infty$
otherwise. Importantly, even for non-convex domains
$\mathbb{A}$, an abstract step function 
allows testing whether an interval $[L, U]$ intersects $a \in
\mathbb{A}$ using at most one membership test and at
most one $\mathrm{step}_{\mathbb{A}}$ evaluation, as we explain next. 

\begin{theorem}[One-Step Interval Intersection]
\label{thm:one_step}
Let $\mathbb{A}$ be an abstract domain over a totally ordered set
$(\mathbb{C}, \le)$ with concretization $\gamma_{\mathbb{A}}$. For any
$a \in \mathbb{A}$ and contiguous interval $[L, U]$ over $\mathbb{C}$:
\[
  [L, U] \cap \gamma_{\mathbb{A}}(a) \neq \emptyset \iff L \in \gamma_{\mathbb{A}}(a) \;\lor\; \mathrm{step}_{\mathbb{A}}(a, L) \le U.
\]
\end{theorem}

\begin{proof}
$(\implies)$ If $[L, U] \cap \gamma_{\mathbb{A}}(a) \neq \emptyset$,
there exists $x \in \gamma_{\mathbb{A}}(a)$ with $L \le x \le U$. If
$x = L$, then $L \in \gamma_{\mathbb{A}}(a)$. If $x > L$,  then
by the corollary of Definition~\ref{def:step}, $L <
\mathrm{step}_{\mathbb{A}}(a, L) \le x
\le U$, so $\mathrm{step}_{\mathbb{A}}(a, L) \le U$. $(\impliedby)$ If
$L \in \gamma_{\mathbb{A}}(a)$, then $L \in [L, U] \cap
\gamma_{\mathbb{A}}(a) \neq \emptyset$. Otherwise, if $x \triangleq
\mathrm{step}_{\mathbb{A}}(a, L) \le U$, then by
Definition~\ref{def:step}, $x \in \gamma_{\mathbb{A}}(a)$ and $L < x
\le U$, so $x \in [L, U] \cap \gamma_{\mathbb{A}}(a) \neq \emptyset$.
\end{proof}

\subsection{A Step Function for the Tnum Domain}
\label{sec:tnum_step}

We now instantiate the abstract step operator for the tristate domain
$\mathbb{A}_{\mathit{tnum}}$, named \texttt{tnum\_step}. A tristate
number $t = (v, m) \in \mathbb{W}_{64} \times \mathbb{W}_{64}$ ($v
\mathbin{\&} m = 0$) has concretization $\gamma_{\mathit{tnum}}(t) =
\{x \in \mathbb{W}_{64} \mid (x \mathbin{\&} \mathord{\sim} m) = v\}$,
with minimum member $t_{\min} \triangleq v$ and maximum member
$t_{\max} \triangleq v \mathbin{|} m$. For an arbitrary machine word
$z \in \mathbb{W}_{64}$, the tristate step operator computes
$\mathtt{tnum\_step}(t, z) \triangleq \min\{w \in
\gamma_{\mathit{tnum}}(t) \mid w > z\}$. We make \texttt{tnum\_step()}
a total function over $\mathbb{W}_{64}$ 
by adding two simplifying boundary conditions:  (1)~If $z \ge
t_{\max}$, no strictly greater element exists in
$\gamma_{\mathit{tnum}}(t)$, and the procedure returns $t_{\max}$.
(2) If $z < t_{\min}$, 
the procedure returns $t_{\min} = v$.
The core problem is then finding the successor of $z$ when $z$ lies
within $t$'s bounding box ($t_{\min} \le z < t_{\max}$). Because
$\gamma_{\mathit{tnum}}(t)$ is a sparse, non-contiguous set of up to
$2^{64}$ integers separated by irregular gaps, finding the next valid
member larger than $z$ may appear to require enumerating bit-patterns
or concrete values. However, as we show next, we can compute
\texttt{tnum\_step($t, z$)} in $O(1)$ time using loop-free bitwise
operations.

\Para{Intuition.} If all the bits of the tnum $t$ were unknown
(\texttt{x}) and free to vary, \texttt{tnum\_step($t, z$)} is simple:
increment $z$ and return $z + 1.$ Hence, the main challenge is the
presence of the fixed bits in $t$. We start by overwriting the
specific bits of $z$ corresponding to the fixed bits of $t$.  We call
the result the {\em projection} of $z$ onto $t$, since it is a member
of $t$.  The projection may not be minimally larger than $z$, or even
larger than $z$, as the step function mandates. Hence we ``fix'' the
projection. At a high level, if the projection is larger than $z$, we
identify the most-significant bit position that makes $z$ smaller than
$v$, and clear out all the free bits in lesser-significant
positions. This makes the result minimally larger than $z$. If the
projection is smaller than $z$, we identify the most-significant bit
position that makes $z$ larger than $v$. We then ``increment'' the
corresponding high-order prefix while leaving the fixed bits
unchanged. We now discuss the details.

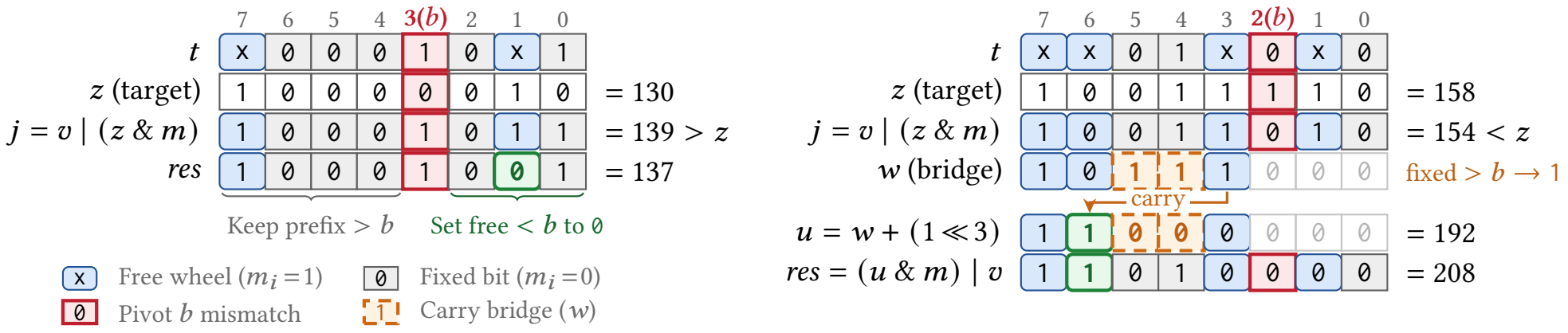
\begin{figure}[t]
\centering
\begin{tikzpicture}[
  font=\scriptsize,
  cell/.style={draw=gray!75!black, minimum width=0.38cm, minimum height=0.30cm, inner sep=0pt, anchor=center, font=\ttfamily\scriptsize},
  weld/.style={cell, fill=gray!13, append after command={
    \pgfextra{
      \fill[gray!65!black] ($(\tikzlastnode.north west)+(0.035,-0.035)$) circle (0.6pt);
      \fill[gray!65!black] ($(\tikzlastnode.south east)+(-0.035,0.035)$) circle (0.6pt);
    }
  }},
  unk/.style={cell, fill=lightblue, draw=cblue!85!black, rounded corners=1.4pt},
  drum/.style={unk, append after command={
    \pgfextra{
      \draw[cblue!60!black, line width=0.35pt] ($(\tikzlastnode.north west)+(0.04,-0.028)$) -- ($(\tikzlastnode.north east)+(-0.04,-0.028)$);
      \draw[cblue!60!black, line width=0.35pt] ($(\tikzlastnode.south west)+(0.04,0.028)$) -- ($(\tikzlastnode.south east)+(-0.04,0.028)$);
    }
  }},
  flip/.style={cell, fill=lightred, draw=cred, thick},
  weldflip/.style={flip, append after command={
    \pgfextra{
      \fill[cred!80!black] ($(\tikzlastnode.north west)+(0.035,-0.035)$) circle (0.6pt);
      \fill[cred!80!black] ($(\tikzlastnode.south east)+(-0.035,0.035)$) circle (0.6pt);
    }
  }},
  bridge/.style={cell, fill=lightorange, draw=corange, densely dashed, thick, text=corange!90!black, font=\ttfamily\scriptsize\bfseries},
  bump/.style={unk, fill=lightgreen, draw=cgreen, thick, text=cgreen!80!black, font=\ttfamily\scriptsize\bfseries},
  dimcell/.style={cell, draw=gray!45, text=gray!60},
  lbl/.style={anchor=east, font=\scriptsize},
  rval/.style={anchor=west, font=\scriptsize}
]
  \begin{scope}[xshift=0cm]
    \node[font=\scriptsize\bfseries, anchor=south west, align=center] at (-1.55, 1.62) {(a) Case 1 ($j > z$)};
    \foreach \i/\b in {0/7, 1/6, 2/5, 3/4, 5/2, 6/1, 7/0} {
      \node[font=\tiny, text=gray!80!black] at (\i*0.38, 1.45) {\b};
    }
    \node[font=\tiny\bfseries, text=cred] at (4*0.38, 1.46) {3($b$)};

    \node[lbl] at (-0.22, 1.18) {$t$};
    \node[drum] at (0*0.38, 1.18) {x}; \node[weld] at (1*0.38, 1.18) {0}; \node[weld] at (2*0.38, 1.18) {0}; \node[weld] at (3*0.38, 1.18) {0};
    \node[weldflip] at (4*0.38, 1.18) {1}; \node[weld] at (5*0.38, 1.18) {0}; \node[drum] at (6*0.38, 1.18) {x}; \node[weld] at (7*0.38, 1.18) {1};

    \node[lbl] at (-0.22, 0.85) {$z$ (target)};
    \node[cell] at (0*0.38, 0.85) {1}; \node[cell] at (1*0.38, 0.85) {0}; \node[cell] at (2*0.38, 0.85) {0}; \node[cell] at (3*0.38, 0.85) {0};
    \node[flip] at (4*0.38, 0.85) {0}; \node[cell] at (5*0.38, 0.85) {0}; \node[cell] at (6*0.38, 0.85) {1}; \node[cell] at (7*0.38, 0.85) {0};
    \node[rval] at (2.88, 0.85) {$= 130$};

    \node[lbl] at (-0.22, 0.52) {$j = v \mathbin{|} (z \mathbin{\&} m)$};
    \node[unk] at (0*0.38, 0.52) {1}; \node[weld] at (1*0.38, 0.52) {0}; \node[weld] at (2*0.38, 0.52) {0}; \node[weld] at (3*0.38, 0.52) {0};
    \node[weldflip] at (4*0.38, 0.52) {1}; \node[weld] at (5*0.38, 0.52) {0}; \node[unk] at (6*0.38, 0.52) {1}; \node[weld] at (7*0.38, 0.52) {1};
    \node[rval] at (2.88, 0.52) {$= 139 > z$};

    \node[lbl] at (-0.22, 0.19) {$\mathit{res}$};
    \node[unk] at (0*0.38, 0.19) {1}; \node[weld] at (1*0.38, 0.19) {0}; \node[weld] at (2*0.38, 0.19) {0}; \node[weld] at (3*0.38, 0.19) {0};
    \node[weldflip] at (4*0.38, 0.19) {1}; \node[weld] at (5*0.38, 0.19) {0}; \node[bump] at (6*0.38, 0.19) {0}; \node[weld] at (7*0.38, 0.19) {1};
    \node[rval] at (2.88, 0.19) {$= 137$};

    \draw[decorate, decoration={brace, mirror, amplitude=1.6pt}, gray!80!black] (-0.17, 0.01) -- (1.31, 0.01) node[midway, below=1.5pt, font=\tiny] {Keep prefix $> b$};
    \draw[decorate, decoration={brace, mirror, amplitude=1.6pt}, cgreen!75!black] (1.73, 0.01) -- (2.83, 0.01) node[midway, below=1.5pt, font=\tiny, text=cgreen!80!black] {Set free $< b$ to \texttt{0}};

    \node[drum, minimum width=0.28cm, minimum height=0.19cm, font=\ttfamily\tiny] at (-1.35, -0.70) {x};
    \node[anchor=west, font=\tiny, text=gray!85!black] at (-1.15, -0.70) {Free wheel  ($m_i\!=\!1$)};
    \node[weld, minimum width=0.28cm, minimum height=0.19cm, font=\ttfamily\tiny] at (1.15, -0.70) {0};
    \node[anchor=west, font=\tiny, text=gray!85!black] at (1.35, -0.70) {Fixed bit ($m_i\!=\!0$)};

    \node[weldflip, minimum width=0.28cm, minimum height=0.19cm, font=\ttfamily\tiny] at (-1.35, -0.98) {0};
    \node[anchor=west, font=\tiny, text=gray!85!black] at (-1.15, -0.98) {Pivot $b$ mismatch};
    \node[bridge, minimum width=0.28cm, minimum height=0.19cm, font=\ttfamily\tiny] at (1.15, -0.98) {1};
    \node[anchor=west, font=\tiny, text=gray!85!black] at (1.35, -0.98) {Carry bridge ($w$)};
  \end{scope}

  \begin{scope}[xshift=6.65cm]
    \node[font=\scriptsize\bfseries, anchor=south west, align=center] at (-1.65, 1.62) {(b) Case 2 ($j \le z$)};
    \foreach \i/\b in {0/7, 1/6, 2/5, 3/4, 4/3, 6/1, 7/0} {
      \node[font=\tiny, text=gray!80!black] at (\i*0.38, 1.45) {\b};
    }
    \node[font=\tiny\bfseries, text=cred] at (5*0.38, 1.46) {2($b$)};

    \node[lbl] at (-0.22, 1.18) {$t$};
    \node[drum] at (0*0.38, 1.18) {x}; \node[drum] at (1*0.38, 1.18) {x}; \node[weld] at (2*0.38, 1.18) {0}; \node[weld] at (3*0.38, 1.18) {1};
    \node[drum] at (4*0.38, 1.18) {x}; \node[weldflip] at (5*0.38, 1.18) {0}; \node[drum] at (6*0.38, 1.18) {x}; \node[weld] at (7*0.38, 1.18) {0};

    \node[lbl] at (-0.22, 0.85) {$z$ (target)};
    \node[cell] at (0*0.38, 0.85) {1}; \node[cell] at (1*0.38, 0.85) {0}; \node[cell] at (2*0.38, 0.85) {0}; \node[cell] at (3*0.38, 0.85) {1};
    \node[cell] at (4*0.38, 0.85) {1}; \node[flip] at (5*0.38, 0.85) {1}; \node[cell] at (6*0.38, 0.85) {1}; \node[cell] at (7*0.38, 0.85) {0};
    \node[rval] at (2.88, 0.85) {$= 158$};

    \node[lbl] at (-0.22, 0.52) {$j = v \mathbin{|} (z \mathbin{\&} m)$};
    \node[unk] at (0*0.38, 0.52) {1}; \node[unk] at (1*0.38, 0.52) {0}; \node[weld] at (2*0.38, 0.52) {0}; \node[weld] at (3*0.38, 0.52) {1};
    \node[unk] at (4*0.38, 0.52) {1}; \node[weldflip] at (5*0.38, 0.52) {0}; \node[unk] at (6*0.38, 0.52) {1}; \node[weld] at (7*0.38, 0.52) {0};
    \node[rval] at (2.88, 0.52) {$= 154 < z$};

    \node[lbl] at (-0.22, 0.19) {$w$ (bridge)};
    \node[unk] at (0*0.38, 0.19) {1}; \node[unk] at (1*0.38, 0.19) {0}; \node[bridge] at (2*0.38, 0.19) {1}; \node[bridge] at (3*0.38, 0.19) {1};
    \node[unk] at (4*0.38, 0.19) {1}; \node[dimcell] at (5*0.38, 0.19) {0}; \node[dimcell] at (6*0.38, 0.19) {0}; \node[dimcell] at (7*0.38, 0.19) {0};
    \node[rval, text=corange!90!black, font=\tiny] at (2.88, 0.19) {fixed $>b \to \mathtt{1}$};

    \draw[-{Stealth[length=3.5pt,width=3pt]}, semithick, corange!95!black]
      (4*0.38, 0.02) -- (4*0.38, -0.07) -- node[midway, fill=white, inner sep=1pt, font=\tiny, text=corange!95!black] {carry} (1*0.38, -0.07) -- (1*0.38, -0.16);

    \node[lbl] at (-0.22, -0.32) {$u = w + (1 \!\ll\! 3)$};
    \node[unk] at (0*0.38, -0.32) {1}; \node[bump] at (1*0.38, -0.32) {1}; \node[bridge] at (2*0.38, -0.32) {0}; \node[bridge] at (3*0.38, -0.32) {0};
    \node[unk] at (4*0.38, -0.32) {0}; \node[dimcell] at (5*0.38, -0.32) {0}; \node[dimcell] at (6*0.38, -0.32) {0}; \node[dimcell] at (7*0.38, -0.32) {0};
    \node[rval] at (2.88, -0.32) {$= 192$};

    \node[lbl] at (-0.22, -0.65) {$\mathit{res} = (u \mathbin{\&} m) \mathbin{|} v$};
    \node[unk] at (0*0.38, -0.65) {1}; \node[bump] at (1*0.38, -0.65) {1}; \node[weld] at (2*0.38, -0.65) {0}; \node[weld] at (3*0.38, -0.65) {1};
    \node[unk] at (4*0.38, -0.65) {0}; \node[weldflip] at (5*0.38, -0.65) {0}; \node[unk] at (6*0.38, -0.65) {0}; \node[weld] at (7*0.38, -0.65) {0};
    \node[rval] at (2.88, -0.65) {$= 208$};
  \end{scope}
\end{tikzpicture}
\vspace{-0.65em}
\caption{\textbf{Binary execution of \texttt{tnum\_step}.}
Blue cells are free bits ($m_i=1$); gray cells are fixed bits
($m_i=0$); red marks pivot $b$. (a)~Case~1 ($j > z$): free bits $<
b$ roll to \texttt{0} (green), $>b$ match $z$. (b)~Case~2 ($j \le z$):
bridge $w$ sets fixed bits $4, 5$ to \texttt{1} (orange) so $+(1 \ll
3)$ ripples across bit~5 into bit~6.}
\label{fig:tnum_step}
\end{figure}

\textbf{Step 1: Greedy projection (candidate $j$).} The natural
starting point is a greedy approximation: copy $z$'s bits directly
into every free bit of $t$ while keeping all known bits fixed to
$v$, producing the initial candidate $j$. By construction, $j \in
\gamma_{\mathit{tnum}}(t)$. Because $j$ and $z$ are \emph{identical}
on every unknown bit position, they can only differ at known bit
positions where $v_i \neq z_i$.

\textbf{Step 2: Identifying the pivot bit $b$.}
Suppose $j \neq z$, and let $b \in \{0, \dots, 62\}$ denote the
0-indexed \emph{most significant bit} where $j$ and $z$ disagree. The
weight of bit $b$ ($2^b$) strictly exceeds the combined sum of all
lower bits ($\sum_{i=0}^{b-1} 2^i = 2^b - 1$). Above bit $b$, $j$ and
$z$ are bit-for-bit identical ($j_{[63:b+1]} = z_{[63:b+1]}$).
Therefore, the numerical ordering between $j$ and $z$, which
determines how $j$ must be adjusted to form the minimal successor of
$z$, is governed entirely by pivot bit $b$. Depending on whether $j >
z$ or $j \le z$, two cases arise:

\textbf{Step 3, Case 1 ($j > z$: 0-to-1 flip at pivot $b$).}
Here $j$ already exceeds $z$ because at the highest disagreement bit
$b$, $z$ has a \texttt{0} while $t$ has a known \texttt{1}. Because
$2^b > \sum_{i=0}^{b-1} 2^i$, that single known \texttt{1} at bit $b$
already forces \emph{any} candidate that matches $z$ above $b$ to be
strictly greater than $z$, regardless of how the free bits below $b$
are set. Conversely, changing any free bit above $b$ from \texttt{1}
to \texttt{0} would make the number strictly smaller than $z$, and
changing a free bit above $b$ from \texttt{0} to \texttt{1} would
make it larger than $j$. Thus, to make our candidate \emph{minimally}
greater than $z$, we keep all free bits above $b$ matched to
$z$ and reset all free bits below $b$ to \texttt{0}.

In Figure~\ref{fig:tnum_step}(a), $t = \mathtt{x00010x1}_2$ and $z =
130 = \mathtt{10000010}_2$. Greedy projection copies $z_7 = \mathtt{1}$
and $z_1 = \mathtt{1}$ into the two `\texttt{x}' positions, yielding
$j = \mathtt{10001011}_2 = 139 > z$. The highest disagreement is at
bit $b = 3$, where $z_3 = \mathtt{0}$ but $t$ has a known \texttt{1}.
Keeping unknown bits above $b$ equal to $z$ (\ie, setting bit~$7$ to
$z_7 = 1$) and clearing unknown bits below $b$ (\ie clearing bit~$1$
to \texttt{0}) yields the exact minimal successor $\mathtt{10001001}_2
= 137$.

\textbf{Step 3, Case 2 ($j \le z$: 1-to-0 flip at pivot $b$, or $j =
  z$: an exact match).}  When $j < z$, at the highest disagreement bit
$b$, $z$ has a \texttt{1} while $t$ has a known
\texttt{0}. Candidate $j$ falls behind $z$ by $2^b$. Even if we set
every free bit below $b$ to \texttt{1}, their total contribution is
at most $\sum_{i=0}^{b-1} 2^i = 2^b - 1 < 2^b$, which can never make
up for this deficit to make it exceed $z$. Consequently, to strictly
exceed $z$, we must ``increment'' the free bits strictly above $b$
(or across all free bit positions if $j = z$). Specifically, we must
find the lowest free bit $c > b$ where $z_c = 0$ and flip it to
\texttt{1}, keep all higher free bits above $c$ matched to $z$, and
reset all free bits below $c$ to \texttt{0} to guarantee
minimality.

\begin{wrapfigure}{R}{0.56\textwidth}
\input{figs/alg_tnum_step.tex}
\end{wrapfigure}

How can we increment the next free bit above $b$ without iterating
over intermediate bits? If we simply added $1 \ll (b + 1)$ to $j$, the
carry would stop at the very first \texttt{0} bit encountered. If that
position corresponds to a known \texttt{0} in $t$, addition would
incorrectly flip that fixed bit to \texttt{1} instead of propagating
past it to reach a higher free bit. To overcome this, we construct a
{\em carry bridge} word $w$: at all bit positions above $b$, we copy $z$'s
bits on the free bits as usual, but force all known bit
positions above $b$ to \texttt{1}. Because a bit pre-loaded with
\texttt{1} propagates an incoming carry ($1 + 1 = 0$ with carry $1$),
adding $1 \ll (b + 1)$ to $w$ (producing sum $u$) ripples the carry
across all intermediate fixed bits, automatically depositing it into
the lowest free bit $c > b$ where $z_c = 0$, while also
clearing all lower bits between $b + 1$ and $c - 1$. Finally,
extracting only the free bits from $u$ and restoring the fixed bits
from $t$ yields the minimal successor.

In Figure~\ref{fig:tnum_step}(b), projecting $z = 158 =
\mathtt{10011110}_2$ onto $t = \mathtt{xx01x0x0}_2$ yields $j =
\mathtt{10011010}_2 = 154 < z$ due to a fixed \texttt{0} at bit $b =
2$. Bridge $w$ copies free bits $\ge 3$ from $z$, clears bits $<
3$, and sets known bits $4$ and $5$ to \texttt{1} ($w =
184$). Adding $1 \ll 3$ ($8$) propagates the carry through bits $3, 4,
5$ into free bit position $6$, yielding $u = 192$. Finally,
restoring fixed bits yields the minimal successor $208$.

\Para{Algorithm.} Algorithm~\ref{alg:tnum_step} summarizes the complete
procedure. The kernel helper \texttt{fls64($p$)} returns the 1-indexed
position $k = b + 1 \in \{1, \dots, 64\}$ of the most significant set
bit in $p$ ($\mathtt{fls64}(0) = 0$), compiling to a single
instruction (\texttt{bsrq} on x86-64, \texttt{clz} on ARM64). Because
the boundary guards on line~1 ensure $t_{\min} \le z < t_{\max}$, $j$
and $z$ can never disagree at the MSB $b = 63$, so $0 \le k \le 63$
always holds and the shifts by $k$ never shift by 64.

\Para{Formal Verification.}  To prove that \texttt{tnum\_step()} is
sound and optimal for all 64-bit inputs, we verified it against a
first-order bit-vector specification in
Equation~(\ref{eq:tnum_step_spec}) below. 

{
\begin{equation}
\begin{aligned}
  &\forall t \in \mathbb{A}_{\mathit{tnum}},\; z, x \in \mathbb{W}_{64} :\; (z < t_{\max}) \land x = \mathtt{tnum\_step}(t, z) \\[-2pt]
  &\quad \implies x \in \gamma_{\mathit{tnum}}(t) \;\land\; x > z \;\land\; \big(\neg\exists y \in \mathbb{W}_{64} : y \in \gamma_{\mathit{tnum}}(t) \land y > z \land y < x\big)
\end{aligned}
\label{eq:tnum_step_spec}
\end{equation}
}

For any $t \in
\mathbb{A}_{\mathit{tnum}}$ and $z \in \mathbb{W}_{64}$ with $z <
t_{\max}$, $x \triangleq \mathtt{tnum\_step}(t, z)$ must be the unique
smallest element in $\gamma_{\mathit{tnum}}(t)$ strictly greater than
$z$. Negating Equation~(\ref{eq:tnum_step_spec}) over $y \in
\mathbb{W}_{64}$ yields a \texttt{QF\_BV} query that Z3~\cite{z3} and
Agni's LLVM IR verifier~\cite{agni-cav23} proved \texttt{unsat} in
seconds. We also have a bitwidth-parametric proof in Lean~4. The step
operator is merged in mainline Linux, and kernel maintainers have
adopted \texttt{tnum\_step} in reduction to detect singleton
intersections~\cite{chaignon2026singleton} and tighten interval
bounds~\cite{alexei-trim-2026}. The community has also used AI coding
agents to synthesize a strength-reduced
implementation~\cite{sun2026tnumstep,sun2027vero}.

  \section{Multi-Domain Set Intersections}
\label{sec:general}

Equipped with domain-specific step functions, we now formulate a
general search procedure to decide joint set intersections across an
arbitrary number of abstract domains (\S\ref{sec:general:algorithm})
and prove sufficient conditions under which multi-domain intersection
runs in $O(1)$ time (\S\ref{sec:general:sufficient}).

\subsection{A General Algorithm for Multi-Domain Set Intersections} 
\label{sec:general:algorithm}
\begin{wrapfigure}[23]{L}{0.51\textwidth}
\begin{algorithm}[H]
\SetAlCapFnt{\small}
\SetAlCapNameFnt{\small}
\SetAlCapHSkip{0pt}
\caption{General Multi-Domain Set Intersection using Abstract Step Functions}
\label{alg:general_step}
\small
\providecommand{\commstyle}[1]{\textcolor{black!55}{\textit{#1}}}
\SetCommentSty{commstyle}
\SetKwComment{tcp}{}{}
\KwIn{Abstract values $a_1 \in \mathbb{A}_1, \dots, a_k \in \mathbb{A}_k$ over totally ordered concrete domain $\mathbb{C}$.} 

\KwOut{%
\begin{tabular}[t]{@{}lll@{}}
\textbf{true}  & if $\bigcap_{i=1}^k \gamma_i(a_i) \neq \emptyset$ & ($CM$) \\
\textbf{false} & if $\bigcap_{i=1}^k \gamma_i(a_i) = \emptyset$ & ($EI$)
\end{tabular}%
} 
$x_c \leftarrow x_{min} \leftarrow \max_{1 \le i \le k} \big( \min(\gamma_i(a_i)) \big)$\;

$x_{max} \leftarrow \min_{1 \le i \le k} \big( \max(\gamma_i(a_i)) \big)$\; 

\While{$x_c \le x_{max}$}
    { 
    $\textnormal{all\_contain} \leftarrow \textbf{true}$\; 
    \For{$i \leftarrow 1$ \KwTo $k$}
        { \If{$x_c \notin \gamma_i(a_i)$}
            { $\textnormal{all\_contain} \leftarrow \textbf{false}$\; 
              $x_c \leftarrow x_c^* \leftarrow \text{step}_i(a_i, x_c)$ \tcp*[r]{update $x_c$}
              \textbf{break}\; 
            } 
        } 
    \If{\textnormal{all\_contain}}
        {\Return \textbf{true} \tcp*[r]{Common member $x_c$}} 
    } 
    \Return \textbf{false} \tcp*[r]{No intersection}
\end{algorithm}

\end{wrapfigure}

Let $\mathbb{A}_1, \dots, \mathbb{A}_k$ be $k$ abstract domains over a
shared, totally ordered concrete set $\mathbb{C}$. Let $a_1 \in
\mathbb{A}_1, \dots, a_k \in \mathbb{A}_k$ be well-formed abstract
values. Given any $a_i \in \mathbb{A}_i$ and any $x \in \mathbb{C}$,
we assume that each domain $\mathbb{A}_i$ provides: (1)
\textbf{Minimum}: $\min(\gamma_i(a_i))$, (2) \textbf{Maximum}:
$\max(\gamma_i(a_i))$, (3) \textbf{Membership}: testing $x \in
\gamma_i(a_i)$, and (4) \textbf{Step}: $\textnormal{step}_i(a_i, x)
\!= \!\min \{ y \! \in \! \gamma_i(a_i) \mid y > x\}$.

Algorithm~\ref{alg:general_step} details our decision procedure, which
operates like the multi-way merge phase of merge-sort.  It initializes
$x_{\min}$ to the maximum of all domain minimums and $x_{\max}$ to the
minimum of all domain maximums (lines 1-2). Since no common element
can lie outside $[x_{\min}, x_{\max}]$, it starts the candidate search
at $x_c \leftarrow x_{\min}$.
If $x_c > x_{\max}$, the domains
are separated by an empty gap and the algorithm returns \texttt{false}
(line~16).
Otherwise, it tests whether $x_c$ belongs to all $k$ abstract values
(line~6). If some $a_i \in \mathbb{A}_i$ does not contain $x_c$, it
advances $x_c$ using $\mathbb{A}_i$'s step function: $x_c \leftarrow
x_c^* \leftarrow \mathrm{step}_i(a_i, x_c)$ (line~8).

Because $\mathrm{step}_i(a_i, x_c)$ returns the smallest member of
$\gamma_i(a_i)$ strictly greater than $x_c$, no value in $[x_c,
x_c^*)$ can belong to $\gamma_i(a_i)$. Hence, advancing $x_c
\leftarrow x_c^*$ is complete (skips no common members) and sound
(strictly increases $x_c$). Stepping repeats until either $x_c$
satisfies all $k$ domains, returning \texttt{true} (line~13), or $x_c
> x_{\max}$, returning \texttt{false} (line~16).

\begin{figure}[t]
\centering
\begin{subfigure}[b]{0.49\linewidth}
\centering
\begin{tikzpicture}[x=0.45cm, y=0.7cm]
    \draw[thick, ->] (-0.3, 0) -- (12.5, 0) node[right, font=\footnotesize\bfseries] {$\mathbb{C}$}; 
    \foreach \n in {0,1,2,3,4,5,6,7,8,9,10,11,12} { 
        \draw (\n, 0.08) -- (\n, -0.08) node[below=1pt, font=\footnotesize] {\n}; 
    }

    \draw[thick, blue!70!black, dashed] (2, 0) -- (2, 2.8) 
        node[above, font=\footnotesize\bfseries, text=blue!70!black] {$x_{\min}$};

    \draw[thick, red!70!black, dashed] (11, 0) -- (11, 2.8) 
        node[above, font=\footnotesize\bfseries, text=red!70!black] {$x_{\max}$};

    \node[anchor=east, font=\footnotesize\bfseries, text=blue!80!black] at (-0.35, 2.4) {$\mathbb{A}_1$}; 
    \draw[dashed, blue!30] (0, 2.4) -- (12, 2.4);
    \fill[blue!80!black] (1, 2.4) circle (2.5pt); 
    \fill[blue!80!black] (5, 2.4) circle (2.5pt); 
    \fill[blue!80!black] (9, 2.4) circle (2.5pt);
    \fill[blue!80!black] (12, 2.4) circle (2.5pt);

    \node[anchor=east, font=\footnotesize\bfseries, text=orange!80!black] at (-0.35, 1.55) {$\mathbb{A}_2$}; 
    \draw[dashed, orange!30] (0, 1.55) -- (12, 1.55); 
    \fill[orange!80!black] (2, 1.55) circle (2.5pt); 
    \fill[orange!80!black] (5, 1.55) circle (2.5pt); 
    \fill[orange!80!black] (7, 1.55) circle (2.5pt);
    \fill[orange!80!black] (9, 1.55) circle (2.5pt);
    \fill[orange!80!black] (11, 1.55) circle (2.5pt);

    \node[anchor=east, font=\footnotesize\bfseries, text=teal!80!black] at (-0.35, 0.7) {$\mathbb{A}_3$}; 
    \draw[dashed, teal!30] (0, 0.7) -- (12, 0.7);
    \fill[teal!80!black] (1, 0.7) circle (2.5pt); 
    \fill[teal!80!black] (3, 0.7) circle (2.5pt); 
    \fill[teal!80!black] (9, 0.7) circle (2.5pt);
    \fill[teal!80!black] (12, 0.7) circle (2.5pt);

    \draw[very thick, green!60!black, fill=green!20, opacity=0.5] (8.6, 0.4)
        rectangle (9.4, 2.7); 
    \node[anchor=south, font=\footnotesize\bfseries] at (9, 2.7) {\color{teal}\ding{51}};

    \node[anchor=east, font=\footnotesize\bfseries, text=black, fill=white, inner sep=1pt] at (1.75, 1.55) {$x_c^0\!=\!2$};
    \draw[->, very thick, purple!80!black] (2, 1.55) to[out=38, in=200] (4.85, 2.38);
    \node[font=\footnotesize, text=purple!85!black, fill=white, inner sep=0.8pt] at (2.5, 2.) {$\text{step}_1(a_1, 2)$};
    \node[anchor=south, font=\footnotesize\bfseries, text=black] at (5.0, 2.5) {$x_c^1\!=\!5$};

    \draw[->, very thick, purple!80!black] (5, 2.4) to[out=-42, in=152] (8.85, 0.75);
    \node[font=\footnotesize, text=purple!85!black, fill=white, inner sep=0.8pt] at (7.0, 1.95) {$\text{step}_3(a_3, 5)$};
    \node[anchor=west, font=\footnotesize\bfseries, text=black, fill=white, inner sep=1pt] at (9.5, 1.05) {$x_c^2\!=\!9$};
\end{tikzpicture}
\caption{Successful Convergence ($R \in CM$)}
\label{fig:step_cm}
\end{subfigure}\hfill
\begin{subfigure}[b]{0.49\linewidth}
\centering
\begin{tikzpicture}[x=0.45cm, y=0.7cm]
    \draw[thick, ->] (-0.3, 0) -- (12.5, 0) node[right, font=\footnotesize\bfseries] {$\mathbb{C}$}; 
    \foreach \n in {0,1,2,3,4,5,6,7,8,9,10,11,12} { 
        \draw (\n, 0.08) -- (\n, -0.08) node[below=1pt, font=\footnotesize] {\n}; 
    }

    \draw[thick, blue!70!black, dashed] (2, 0) -- (2, 2.8) 
        node[above, font=\footnotesize\bfseries, text=blue!70!black] {$x_{\min}$};

    \draw[thick, red!70!black, dashed] (7, 0) -- (7, 2.8) 
        node[above, font=\footnotesize\bfseries, text=red!70!black] {$x_{\max}$};

    \node[anchor=east, font=\footnotesize\bfseries, text=blue!80!black] at (-0.35, 2.4) {$\mathbb{A}_1$}; 
    \draw[dashed, blue!30] (0, 2.4) -- (12, 2.4);
    \fill[blue!80!black] (1, 2.4) circle (2.5pt); 
    \fill[blue!80!black] (5, 2.4) circle (2.5pt); 
    \fill[blue!80!black] (9, 2.4) circle (2.5pt);
    \fill[blue!80!black] (12, 2.4) circle (2.5pt);

    \node[anchor=east, font=\footnotesize\bfseries, text=orange!80!black] at (-0.35, 1.55) {$\mathbb{A}_2$}; 
    \draw[dashed, orange!30] (0, 1.55) -- (12, 1.55); 
    \fill[orange!80!black] (2, 1.55) circle (2.5pt); 
    \fill[orange!80!black] (5, 1.55) circle (2.5pt); 
    \fill[orange!80!black] (7, 1.55) circle (2.5pt);

    \node[anchor=east, font=\footnotesize\bfseries, text=teal!80!black] at (-0.35, 0.7) {$\mathbb{A}_3$}; 
    \draw[dashed, teal!30] (0, 0.7) -- (12, 0.7);
    \fill[teal!80!black] (1, 0.7) circle (2.5pt); 
    \fill[teal!80!black] (3, 0.7) circle (2.5pt); 
    \fill[teal!80!black] (9, 0.7) circle (2.5pt);
    \fill[teal!80!black] (12, 0.7) circle (2.5pt);

    \node[anchor=east, font=\footnotesize\bfseries, text=black, fill=white, inner sep=1pt] at (1.75, 1.55) {$x_c^0\!=\!2$};
    \draw[->, very thick, purple!80!black] (2, 1.55) to[out=38, in=200] (4.85, 2.38); 
    \node[font=\footnotesize, text=purple!85!black, fill=white, inner sep=0.8pt] at (2.5, 2.) {$\text{step}_1(a_1, 2)$};
    \node[anchor=south, font=\footnotesize\bfseries, text=black] at (5.0, 2.5) {$x_c^1\!=\!5$};

    \draw[->, very thick, purple!80!black] (5, 2.4) to[out=-42, in=152] (8.85, 0.75); 
    \node[font=\footnotesize, text=purple!85!black, fill=white, inner sep=0.8pt] at (7.0, 1.95) {$\text{step}_3(a_3, 5)$};
    \node[anchor=west, font=\footnotesize\bfseries, text=black, fill=white, inner sep=1pt] at (9.3, 1.05) {$x_c^2\!=\!9$};

    \node[anchor=south, font=\footnotesize\bfseries] at (9, 2.7) {\color{red}\ding{55}};
\end{tikzpicture}
\caption{Unsuccessful Termination ($R \in EI$)}
\label{fig:step_ei}
\end{subfigure}
\caption{\textbf{Illustration of the General Multi-Domain Set Intersection Algorithm.}
Candidate $x_c$ values hop across domains.
In (a), stepping converges to a common concrete member $9$.
In (b), stepping past $5$ produces candidate $x_c^2=9 > x_{max} = 7$, proving an empty intersection.}
\label{fig:general_step}
\end{figure}
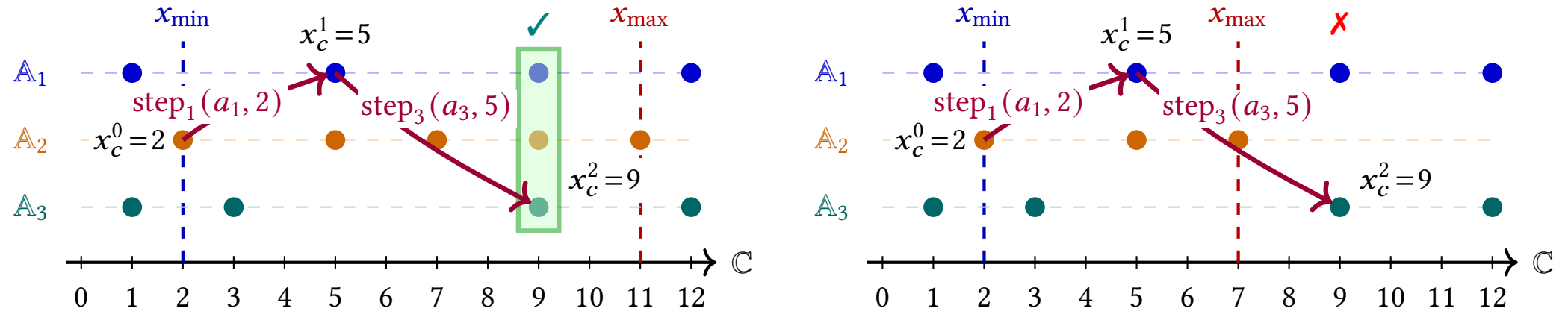

\Para{Visualizing the stepping dynamics.}
Figure~\ref{fig:general_step} illustrates
Algorithm~\ref{alg:general_step} on three domains $\mathbb{A}_1,
\mathbb{A}_2, \mathbb{A}_3$. In Figure~\ref{fig:general_step}(a), $a_1
= \{1, 5, 9, 12\}$, $a_2 = \{2, 5, 7, 9, 11\}$, and $a_3 = \{1, 3, 9,
12\}$, giving $x_c^0 = \max(1, 2, 1) = 2$ and $x_{\max} = \min(12, 11,
12) = 11$. Since $2 \notin a_1$, stepping in $\mathbb{A}_1$ yields
$x_c^1 = \mathrm{step}_1(a_1, 2) = 5$. At $x_c^1 = 5$, $5 \in a_1$ and
$5 \in a_2$, but $5 \notin a_3$, so stepping in $\mathbb{A}_3$ yields
$x_c^2 = \mathrm{step}_3(a_3, 5) = 9$, which belongs to all three
domains, terminating the algorithm with \texttt{true} ($R \in CM$). In
Figure~\ref{fig:general_step}(b), $a_2 = \{2, 5, 7\}$, which lowers
$x_{\max}$ to $\min(12, 7, 12) = 7$; After advancing to $x_c^1 = 5$,
stepping in $\mathbb{A}_3$ yields $x_c^2 = 9 > x_{\max} = 7$,
terminating the algorithm immediately with \texttt{false} ($R \in
EI$).

\Para{Algorithmic Complexity and Termination.}  The number of
iterations in Algorithm~\ref{alg:general_step} is governed by the
number of discrete gaps or \emph{discontinuities} across the domains.
Formally, a discontinuity in $a_i \in \mathbb{A}_i$ is a gap between
consecutive members $u, v \in \gamma_i(a_i)$, \eg
in $\mathbb{A}_{\mathit{u64}}$,  $v > u + 1$.

If $a_i$ has $\delta_i$ discontinuities, $x_c \in [x_{\min},
  x_{\max}]$ can step across gaps in $a_i$ at most $\delta_i$
times. Moreover, because $x_c \leftarrow \mathrm{step}_i(a_i, x_c)$
lands inside $\gamma_i(a_i)$, $\mathbb{A}_i$ cannot step again until
another domain $\mathbb{A}_j \neq \mathbb{A}_i$ steps across one of
its own gaps, bounding total iterations by $O(1 + \sum_{j \neq *}
\delta_j)$ where $\mathbb{A}^*$ is the domain with the most
discontinuities. While intervals have $\delta = 0$, tnums can have
$O(2^w)$ discontinuities over $\mathbb{W}_w$, so Algorithm~2 can take
exponential time only when two or more domains have unbounded
discontinuities.

\subsection{Sufficient Conditions for Constant Time Multi Domain Intersection}
\label{sec:general:sufficient}

We now identify conditions under which multi-domain intersection can
be done in $O(1)$ time.

\begin{theorem}[Sufficient Constant-Time Intersection Theorem]
\label{thm:sufficient}
\label{thm:o1_sufficient}
Let $\mathbb{A}_1, \dots, \mathbb{A}_k$ be $k = O(1)$ abstract domains
over a totally ordered set $\mathbb{C}$. Deciding $\bigcap_{i=1}^k
\gamma_i(a_i) \neq \emptyset$ takes $O(1)$ time if:
\begin{enumerate}[leftmargin=*,noitemsep]
\item[(i)] \textbf{Bounded Discontinuities:} at most one domain
$\mathbb{A}^* \in \{\mathbb{A}_1, \dots, \mathbb{A}_k\}$ has a
non-constant number of discontinuities, while every other domain
$\mathbb{A}_j \neq \mathbb{A}^*$ has at most $C_j = O(1)$
discontinuities;

\item[(ii)] \textbf{Constant-Time Membership:} testing $x \in
\gamma_i(a_i)$ takes $O(1)$ time for each $\mathbb{A}_i$; and

\item[(iii)] \textbf{Constant-Time Step:} computing
$\mathrm{step}_i(a_i, x)$ takes $O(1)$ time for each $\mathbb{A}_i$.
\end{enumerate}
\end{theorem}

\begin{proof}
In Algorithm~\ref{alg:general_step}, since $x_c \ge x_{\min} \ge
\min_j(a_j)$, each step by any $\mathbb{A}_j \neq \mathbb{A}^*$ jumps
across one of $a_j$'s $\le C_j$ internal discontinuities, so all
$\mathbb{A}_j \neq \mathbb{A}^*$ combined execute at most $S \le
\sum_{j \neq *} C_j = O(1)$ steps. Because $x_c \leftarrow \mathrm{step}_*(a^*, x_c)$ places $x_c \in
\gamma_*(a^*)$, $\mathbb{A}^*$ cannot step again until some $\mathbb{A}_j
\neq \mathbb{A}^*$ advances $x_c$; thus $\mathbb{A}^*$ steps at most $S +
1 = O(1)$ times. Algorithm~\ref{alg:general_step} therefore executes at
most $2S + 1$ steps and terminates in at most $2S + 2 = O(1)$ iterations,
each taking $O(k) = O(1)$ time by (ii) and (iii). (When the $C_j + 1$ contiguous intervals of each $\mathbb{A}_j
\neq \mathbb{A}^*$ have explicit endpoints, as in \S\ref{sec:ebpf},
intersecting them yields $M \le \prod_{j \neq *} (C_j + 1) = O(1)$
intervals $[l_p, u_p]$, each testable against $\mathbb{A}^*$ in one
step via Theorem~\ref{thm:one_step}.)
\end{proof}

Theorem~\ref{thm:sufficient} explains why five-domain intersection can
be made $O(1)$ in the eBPF verifier: unsigned intervals
($\mathbb{A}_{\text{u64}}, \mathbb{A}_{\text{u32}}$) have $0$
discontinuities, signed intervals ($\mathbb{A}_{\text{s64}},
\mathbb{A}_{\text{s32}}$) have at most $1$ discontinuity (at the sign
boundary), and any 64-bit interval projects onto $\mathbb{W}_{32}$ as
at most $3$ contiguous slices (\S\ref{sec:discontinuities}), leaving
$\mathbb{A}_{\text{tnum}}$ as the only domain with
an exponential number of gaps.

  \section{All-Domain Abstract Intersection in eBPF}
\label{sec:ebpf}

In the eBPF verifier, registers store 64-bit words $x \in
\mathbb{W}_{64}$, and the five domains in $\mathbb{P}$
(\S\ref{sec:background}) interpret these 64 bits through different
lenses: as unsigned 64-bit integers ($\mathbb{A}_{\text{u64}},
\mathbb{A}_{\text{tnum}}$), two's-complement signed 64-bit integers
($\mathbb{A}_{\text{s64}}$), or lower 32-bit sub-register slices
($\mathbb{A}_{\text{u32}}, \mathbb{A}_{\text{s32}}$).

\subsection{eBPF Domains: Discontinuities and Periodicity}
\label{sec:discontinuities}

While $a_{\text{u64}} \in \mathbb{A}_{\text{u64}}$ concretizes to a
single contiguous interval ($\delta = 0$), the remaining four domains
exhibit discontinuities and periodic gaps when mapped to
$\mathbb{W}_{64}$.

\Para{Signed 64-bit Interval Domain ($\mathbb{A}_{\text{s64}}$)}
constrains $\text{s64}(x) \in [-2^{63}, 2^{63} - 1]$ to $[s64_{\min},
  s64_{\max}]$.
When $s64_{\min} \ge 0$ or $s64_{\max} < 0$, it maps
to a single contiguous unsigned interval in $\mathbb{W}_{64}$ ($\delta
= 0$).
When $[s64_{\min}, s64_{\max}]$ crosses zero ($s64_{\min} < 0
\le s64_{\max}$), negative values map to the upper end of
$\mathbb{W}_{64}$, splitting $\gamma_{\text{s64}}$ into two disjoint
unsigned intervals, $\gamma_{s64}([s64_{\min}, s64_{\max}]) = [0,
  s64_{\max}] \cup [s64_{\min} + 2^{64}, 2^{64}-1]$.
Suppose $a_{s64} = [-1, 1]$. In two's complement, $-1$ is
$2^{64}-1$. The valid unsigned 64-bit register words are $\{0, 1\}
\cup \{2^{64}-1\}$, which introduces one discontinuity ($\delta=1$).

\Para{32-bit sub-register interval domains
  ($\mathbb{A}_{\text{u32}}, \mathbb{A}_{\text{s32}}$)} constrain only
the lower 32 bits $x_{[31:0]}$, leaving the upper 32-bit prefix $k =
x_{[63:32]} \in \mathbb{W}_{32}$ unconstrained. Writing $x = k \cdot
2^{32} + x_{[31:0]}$, let $S_k \triangleq \{x \in \mathbb{W}_{64} \mid
x_{[63:32]} = k\}$ denote the $k$-th \textbf{32-bit subspace}. Because
$\mathbb{A}_{\text{u32}}$ places no constraint on $k$, $[u32_{\min},
  u32_{\max}]$ repeats across all $2^{32}$ subspaces $S_k$. A 32-bit
interval is repeated across every subspace $S_k$ at multiples of the
form $k \cdot 2^{32}$, \ie, $\gamma_{\text{u32}}([u32_{\min},
  u32_{\max}]) = \bigcup_{k=0}^{2^{32}-1} \big[(k \cdot 2^{32}) +
  u32_{\min},\; (k \cdot 2^{32}) + u32_{\max}\big]$.

For example, if $a_{u32} = [5, 10]$, valid 64-bit values include $[5,
  10]$ (for $k=0$), $[2^{32}+5, 2^{32}+10]$ (for $k=1$), and so on,
creating $\delta = 2^{32} - 1$ discontinuities of length
$2^{32}-6$.
Signed 32-bit intervals $a_{\text{s32}} \in \mathbb{A}_{\text{s32}}$
follow the same $2^{32}$-periodic structure, splitting each 32-bit
slice into two unsigned sub-intervals when $s32_{\min} < 0 \le
s32_{\max}$.

\Para{Tristate Numbers ($\mathbb{A}_{\text{tnum}}$)} constrains
individual bit positions via $t = (v, m)$, inducing up to $\delta =
2^{63} - 1$ discontinuities across
$\mathbb{W}_{64}$ (e.g., $v = 0, m = \mathtt{0xff...fe}$
restricts $x$ to even integers).

\begin{figure}[t]
\centering
\begin{minipage}[c]{0.56\linewidth}
\centering
\begin{tikzpicture}[x=0.64cm, y=0.95cm]
    \draw[thick, ->] (0.2, 2.2) -- (11.0, 2.2) node[right, font=\small\bfseries] {$\mathbb{W}_{64}$};

    \draw[dashed, black!40] (1.0, 1.8) -- (1.0, 2.65);
    \node[above=1pt, font=\small, text=black!70] at (1.0, 2.65) {$x_{\text{hi}}\!\ll\!32$};

    \draw[dashed, black!40] (4.2, 1.8) -- (4.2, 2.65);
    \node[above=1pt, font=\small, text=black!70] at (4.2, 2.65) {$(x_{\text{hi}}\!+\!1)\!\ll\!32$};

    \draw[dashed, black!40] (7.2, 1.8) -- (7.2, 2.65);
    \node[above=1pt, font=\small, text=black!70] at (7.2, 2.65) {$y_{\text{hi}}\!\ll\!32$};

    \draw[dashed, black!40] (10.5, 1.8) -- (10.5, 2.65);
    \node[above=1pt, font=\small, text=black!70] at (10.5, 2.65) {$(y_{\text{hi}}\!+\!1)\!\ll\!32$};

    \draw[line width=3.5pt, orange!85!black] (1.5, 2.2) -- (4.2, 2.2);
    \node[above=2pt, font=\small\bfseries, text=orange!90!black] at (2.85, 2.2) {$I_{\text{left}}$};
    \fill[blue!90!black] (1.5, 2.2) circle (2.5pt) node[below=3pt, font=\small\bfseries] {$x$};

    \draw[line width=3.5pt, purple!80!black] (4.2, 2.2) -- (7.2, 2.2);
    \node[above=2pt, font=\small\bfseries, text=purple!90!black] at (5.7, 2.2) {$I_{\text{mid}}$};

    \draw[line width=3.5pt, teal!80!black] (7.2, 2.2) -- (8.5, 2.2);
    \node[above=2pt, font=\small\bfseries, text=teal!90!black] at (8.05, 2.2) {$I_{\text{right}}$};
    \fill[blue!90!black] (8.5, 2.2) circle (2.5pt) node[below=3pt, font=\small\bfseries] {$y$};

    \draw[dotted, thick, orange!85!black] (1.5, 2.05) -- (4.6, 1.25);
    \draw[dotted, thick, orange!85!black] (4.2, 2.05) -- (8.1, 1.25);

    \draw[dotted, thick, purple!80!black] (4.2, 2.05) -- (3.1, 0.85);
    \draw[dotted, thick, purple!80!black] (7.2, 2.05) -- (8.1, 0.85);

    \draw[dotted, thick, teal!80!black] (7.2, 2.05) -- (3.1, 0.45);
    \draw[dotted, thick, teal!80!black] (8.5, 2.05) -- (6.4, 0.45);

    \fill[red!12, opacity=0.6] (4.0, 0.0) rectangle (7.1, 1.45);
    \draw[dashed, red!60] (4.0, 0.0) -- (4.0, 1.45);
    \draw[dashed, red!60] (7.1, 0.0) -- (7.1, 1.45);

    \draw[line width=3.5pt, orange!85!black] (4.6, 1.25) -- (8.1, 1.25);
    \node[right=6pt, font=\small\bfseries, text=orange!90!black] at (8.1, 1.25) {$(I_{\text{left}})_{[31:0]}$};

    \draw[line width=3.5pt, purple!80!black] (3.1, 0.85) -- (8.1, 0.85);
    \node[right=6pt, font=\small\bfseries, text=purple!90!black] at (8.1, 0.85) {$(I_{\text{mid}})_{[31:0]}$};

    \draw[line width=3.5pt, teal!80!black] (3.1, 0.45) -- (6.4, 0.45);
    \node[right=6pt, font=\small\bfseries, text=teal!90!black] at (8.1, 0.45) {$(I_{\text{right}})_{[31:0]}$};

    \draw[thick, ->] (0.2, 0.0) -- (11.0, 0.0) node[right, font=\small\bfseries] {$\mathbb{W}_{32}$};
    \draw (3.1, -0.1) -- (3.1, 0.1) node[below=3pt, font=\small] {$0$};
    \draw (8.1, -0.1) -- (8.1, 0.1) node[below=3pt, font=\small] {$2^{32}\!-\!1$};

    \draw[line width=3.5pt, red!75!black] (4.0, 0.0) -- (7.1, 0.0);
    \fill[red!75!black] (4.0, 0.0) circle (2.5pt);
    \fill[red!75!black] (7.1, 0.0) circle (2.5pt);
    \node[below=3pt, font=\small\bfseries, text=red!75!black] at (5.55, -0.05) {Target $I_{32}$};
\end{tikzpicture}
\end{minipage}\hfill
\begin{minipage}[c]{0.41\linewidth}
\caption{\textbf{Stage 2: Projecting a 64-bit Interval across Subspaces into 32-bit Space.}
A 64-bit interval $[x, y]$ is partitioned across 32-bit subspaces into at most three segments:
Left Edge ($I_{\text{left}}$), Middle Full Blocks ($I_{\text{mid}}$), and
Right Edge ($I_{\text{right}}$). Projecting each segment onto its lower 32 bits
$(\cdot)_{[31:0]}$ yields candidate 32-bit intervals in $\mathbb{W}_{32}$, which
are intersected with the target 32-bit bounds $I_{32}$ (red shaded band).}
\label{fig:subspace_decomp}
\end{minipage}
\end{figure}
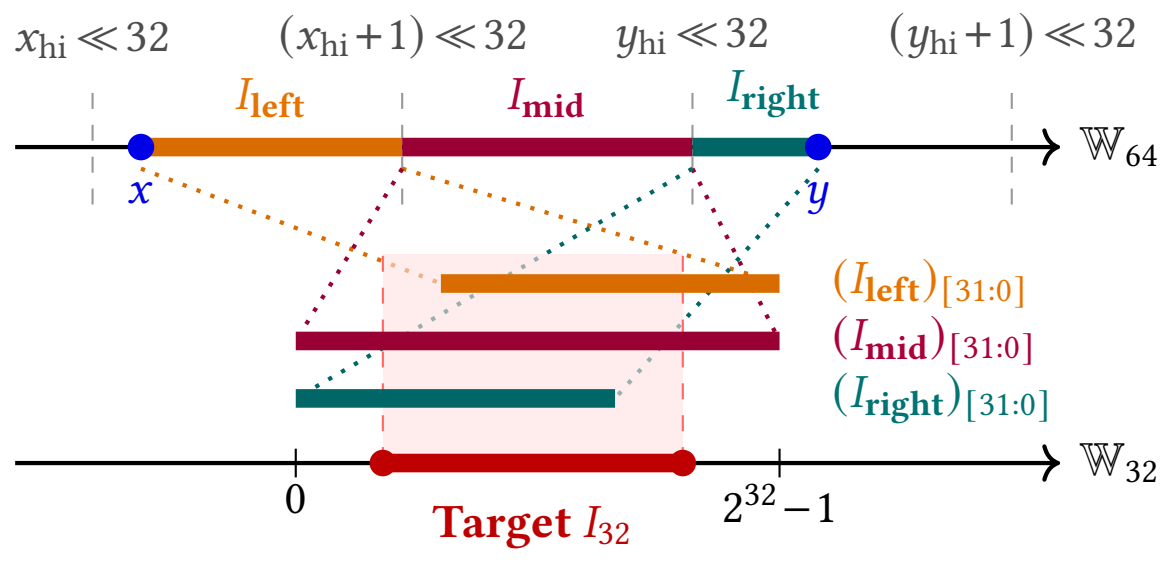

\Para{eBPF domains satisfy Theorem~\ref{thm:sufficient} via subspace
  projection.}
All five eBPF domains satisfy conditions~(ii) and~(iii) of
Theorem~\ref{thm:sufficient}: membership testing takes $O(1)$ bitwise or
comparison operations, intervals have trivial $O(1)$ step functions, and
\texttt{tnum\_step()} (\S\ref{sec:tnum_step}) steps over
$\mathbb{A}_{\text{tnum}}$ in $O(1)$ time.
The central challenge is satisfying condition~(i) (bounded
discontinuities). In the 64-bit space $\mathbb{W}_{64}$, the 64-bit
interval domains $\mathbb{A}_{\text{u64}}$ and $\mathbb{A}_{\text{s64}}$
have $O(1)$ discontinuities ($0$ and at most $1$, respectively). In
contrast, because $\mathbb{A}_{\text{u32}}$ and $\mathbb{A}_{\text{s32}}$
constrain only the lower 32 bits $z_{[31:0]}$ while leaving the upper 32
bits $k = z_{[63:32]}$ unconstrained, every increment of $k$ from $0$ to
$2^{32}-1$ repeats the 32-bit interval across $\mathbb{W}_{64}$. If we
ran Algorithm~\ref{alg:general_step} directly in $\mathbb{W}_{64}$, a
search between $\mathbb{A}_{\text{u32}}$ and $\mathbb{A}_{\text{tnum}}$
could ``ping-pong'' across all $2^{32}$ periodic gaps, requiring billions
of steps.
Crucially, however, this repetition is not irregular: it is strictly
aligned to 32-bit subspace boundaries $S_k = \{z \in \mathbb{W}_{64} \mid
z_{[63:32]} = k\}$. Rather than expanding 32-bit intervals upward into
$2^{32}$ slices in $\mathbb{W}_{64}$, we project 64-bit intervals
downward into the 32-bit space $\mathbb{W}_{32}$, where
$\mathbb{A}_{\text{u32}}$ has $0$ discontinuities and
$\mathbb{A}_{\text{s32}}$ has at most $1$.
For any 64-bit interval $[x, y] \subseteq \mathbb{W}_{64}$ (with upper
32-bit prefixes $x_{\mathrm{hi}} = x_{[63:32]}$ and $y_{\mathrm{hi}} =
y_{[63:32]}$), as a 64-bit counter increments from $x$ to $y$, the set of
lower 32-bit values $\{z_{[31:0]} \mid z \in [x, y]\} \subseteq
\mathbb{W}_{32}$ partitions into at most \textbf{three} contiguous 32-bit
slices (Figure~\ref{fig:subspace_decomp}):
\begin{enumerate}[leftmargin=*,noitemsep]
\item \textbf{Single subspace ($x_{\mathrm{hi}} = y_{\mathrm{hi}}$)}.
  The entire interval $[x, y]$ lies within a single 32-bit subspace
  $S_{x_{\mathrm{hi}}}$ (fixed prefix $k = x_{\mathrm{hi}}$) and projects
  to the single 32-bit interval $[x_{[31:0]}, y_{[31:0]}]$.
\item \textbf{Multiple subspaces ($x_{\mathrm{hi}} < y_{\mathrm{hi}}$)}.
  The interval $[x, y]$ crosses one or more 32-bit subspace boundaries,
  decomposing into:
  \begin{itemize}[leftmargin=*,noitemsep]
  \item \emph{Left partial block ($k = x_{\mathrm{hi}}$)}: $[x, y]$
    starts partway through $S_{x_{\mathrm{hi}}}$ and runs to the top of
    the block, covering lower 32-bit values $[x_{[31:0]}, 2^{32} - 1]$;
  \item \emph{All middle full blocks ($x_{\mathrm{hi}} < k <
    y_{\mathrm{hi}}$)}: Because $[x, y]$ starts before $S_k$ and ends
    after $S_k$, it sweeps through \emph{every} possible lower 32-bit
    value $[0, 2^{32}-1]$ in each intermediate block. Since every
    intermediate block shares the identical lower 32-bit range $[0,
    2^{32}-1]$, up to $2^{32}-2$ middle blocks collapse without loss of
    precision into a \textbf{single} candidate pairing prefix range $k
    \in [x_{\mathrm{hi}} + 1, y_{\mathrm{hi}} - 1]$ with lower 32-bit
    interval $[0, 2^{32} - 1]$;
  \item \emph{Right partial block ($k = y_{\mathrm{hi}}$)}: $[x, y]$
    enters $S_{y_{\mathrm{hi}}}$ at $0$ and stops at $y_{[31:0]}$,
    covering lower 32-bit values $[0, y_{[31:0]}]$.
  \end{itemize}
\end{enumerate}
Consequently, the joint intersection of all four interval domains
decomposes into at most $\mathbf{12}$ exact candidate boxes:
\begin{enumerate}[label=(\roman*),leftmargin=*,noitemsep]
\item Intersecting the 64-bit domains $\mathbb{A}_{\text{u64}} \cap
  \mathbb{A}_{\text{s64}}$ yields at most $\mathbf{2}$ contiguous 64-bit
  intervals in $\mathbb{W}_{64}$ (two when $\mathbb{A}_{\text{s64}}$
  wraps across zero);
\item Decomposing each 64-bit interval into its left, middle, and right
  32-bit subspaces produces at most $2 \times 3 = \mathbf{6}$ projected
  slices, each consisting of a contiguous upper-32-bit prefix range and a
  contiguous lower-32-bit range;
\item Intersecting the lower-32-bit range of each slice with the 32-bit
  domains $\mathbb{A}_{\text{u32}} \cap \mathbb{A}_{\text{s32}}$ (at most
  $\mathbf{2}$ contiguous intervals in $\mathbb{W}_{32}$) yields at most
  $6 \times 2 = \mathbf{12}$ candidate boxes.
\end{enumerate}
Finally, because a tristate bitmask $t \in \mathbb{A}_{\text{tnum}}$
imposes independent bitwise constraints on the upper 32 bits
($t_{\mathrm{hi}}$) and lower 32 bits ($t_{\mathrm{lo}}$) with no carries
between bit positions, each of the $\le 12$ candidate boxes can be tested
against $\mathbb{A}_{\text{tnum}}$ in $O(1)$ time using at most two calls
to \texttt{tnum\_step()} (one on the upper-32-bit prefix interval against
$t_{\mathrm{hi}}$, and one on the lower-32-bit interval against
$t_{\mathrm{lo}}$), guaranteeing a sound, complete, and $O(1)$ decision
procedure.

\subsection{Sound and Complete Allwise Intersection in eBPF}
\label{sec:three_stage}

We now present our complete, constant-time allwise intersection
decision procedure (\texttt{reg\_bounds\_intersect()} in the Linux
kernel). The algorithm organizes the five-domain intersection query
into a three-stage pipeline that exploits the 32-bit subspace
periodicity from \S\ref{sec:discontinuities}.

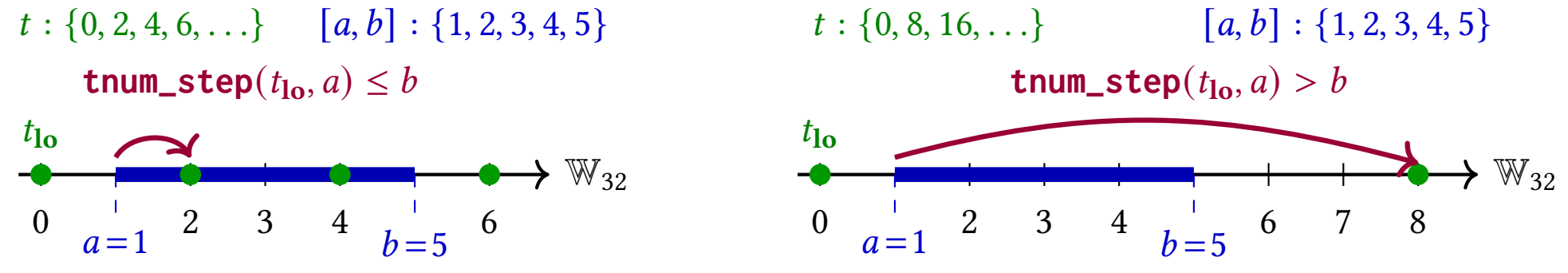
\begin{figure}[t]
\centering
\begin{subfigure}[b]{0.46\linewidth}
\centering
\begin{tikzpicture}[x=0.62cm, y=0.95cm]
    \draw[thick, ->] (-0.3, 0) -- (6.8, 0) node[right, font=\small\bfseries] {$\mathbb{W}_{32}$};
    \foreach \n in {0,2,3,4,6} {
        \draw (\n, 0.1) -- (\n, -0.1) node[below=2pt, font=\small] {\n};
    }

    \draw[line width=3.5pt, blue!70!black] (1, 0) -- (5, 0);
    \draw[dashed, blue!80!black] (1, 0) -- (1, -0.35) node[below, font=\small\bfseries] {$a\!=\!1$};
    \draw[dashed, blue!80!black] (5, 0) -- (5, -0.35) node[below, font=\small\bfseries] {$b\!=\!5$};

    \foreach \n in {0,2,4,6} {
        \fill[green!60!black] (\n, 0) circle (2.5pt);
    }
    \node[anchor=south, font=\small\bfseries, text=green!50!black] at (0, 0.1) {$t_{\text{lo}}$};

    \draw[->, very thick, purple!80!black] (1, 0.15) to[out=60, in=120] (2, 0.15);
    \node[font=\small\bfseries, text=purple!80!black] at (2.8, 0.8) {$\text{\code{tnum\_step}}(t_{\text{lo}}, a) \le b$};

    \node[anchor=west, font=\small\bfseries, text=green!50!black] at (-0.5, 1.3) {$t: \{0, 2, 4, 6, \ldots\}$};
    \node[anchor=east, font=\small\bfseries, text=blue!80!black] at (7.8, 1.3) {$[a, b]: \{1, 2, 3, 4, 5\}$};
\end{tikzpicture}
\caption{Intersection Found $\text{\code{tnum\_step}}(t_{\text{lo}}, a) \le b$}
\label{fig:tnum_step_cm}
\end{subfigure}\hfill
\begin{subfigure}[b]{0.52\linewidth}
\centering
\begin{tikzpicture}[x=0.62cm, y=0.95cm]
    \draw[thick, ->] (-0.3, 0) -- (8.8, 0) node[right, font=\small\bfseries] {$\mathbb{W}_{32}$};
    \foreach \n in {0,2,3,4,6,7,8} {
        \draw (\n, 0.1) -- (\n, -0.1) node[below=2pt, font=\small] {\n};
    }

    \draw[line width=3.5pt, blue!70!black] (1, 0) -- (5, 0);
    \draw[dashed, blue!80!black] (1, 0) -- (1, -0.35) node[below, font=\small\bfseries] {$a\!=\!1$};
    \draw[dashed, blue!80!black] (5, 0) -- (5, -0.35) node[below, font=\small\bfseries] {$b\!=\!5$};

    \fill[green!60!black] (0, 0) circle (2.5pt);
    \fill[green!60!black] (8, 0) circle (2.5pt);
    \node[anchor=south, font=\small\bfseries, text=green!50!black] at (0, 0.1) {$t_{\text{lo}}$};

    \draw[->, very thick, purple!80!black] (1, 0.15) to[out=15, in=165] (8, 0.1);
    \node[font=\small\bfseries, text=purple!80!black] at (4.8, 0.8) {$\text{\code{tnum\_step}}(t_{\text{lo}}, a) > b$};

    \node[anchor=west, font=\small\bfseries, text=green!50!black] at (-0.3, 1.3) {$t: \{0, 8, 16, \ldots\}$};
    \node[anchor=east, font=\small\bfseries, text=blue!80!black] at (9.2, 1.3) {$[a, b]: \{1, 2, 3, 4, 5\}$};
\end{tikzpicture}
\caption{Empty Intersection Detected ($y^* > b$)}
\label{fig:tnum_step_ei}
\end{subfigure}
\caption{\textbf{Stage 3: Interval Intersection via Tnum Stepping
    (Theorem~\ref{thm:one_step}).} In (a), candidate interval $[1, 5]$
  intersects $t$ (even numbers): evaluating
  $\text{\code{tnum\_step}}(t_{\text{lo}}, 1)$ yields $2 \le 5$,
  proving the intersection is non-empty. In (b), $t$ contains only
  multiples of 8: $\text{\code{tnum\_step}}(t_{\text{lo}}, 1)$ jumps
  directly to $8 > 5$; $[1, 5]$ lies entirely within an empty gap,
  proving the intersection is empty.}
\label{fig:tnum_step_fig}
\end{figure}

\Para{Stage 1: 64-Bit and 32-Bit interval intersection.}
The algorithm first intersects the 64-bit intervals $u64 =
[u64_{\min}, u64_{\max}]$ and $s64 = [s64_{\min}, s64_{\max}]$ in
$\mathbb{W}_{64}$, and the 32-bit sub-register intervals $u32 =
[u32_{\min}, u32_{\max}]$ and $s32 = [s32_{\min}, s32_{\max}]$ in
$\mathbb{W}_{32}$.
When $s64$ does not cross zero ($s64_{\min} \ge 0$ or $s64_{\max} <
0$), shifting negative bounds by $+2^{64}$ maps $s64$ to a single
unsigned interval $[s64'_{\min}, s64'_{\max}] \subseteq
\mathbb{W}_{64}$, so $u64 \cap s64$ forms at most one contiguous
unsigned interval $[\max(u64_{\min}, s64'_{\min}), \min(u64_{\max},
  s64'_{\max})]$.
When $s64$ crosses zero ($s64_{\min} < 0 \le
s64_{\max}$), it splits into $s64_0 = [0, s64_{\max}]$ and $s64_1 =
[s64_{\min} + 2^{64}, 2^{64} - 1]$ (\S\ref{sec:background}), and
intersecting both pieces with $u64$ produces a set $\mathcal{I}_{64}$
of at most two contiguous 64-bit unsigned intervals. Intersecting $u32
\cap s32$ in $\mathbb{W}_{32}$ (shifting negative bounds by $+2^{32}$)
analogously produces a set $\mathcal{I}_{32}$ of at most two target
32-bit unsigned intervals ($s32_0 = [0, s32_{\max}]$ and $s32_1 =
[s32_{\min} + 2^{32}, 2^{32} - 1]$):

{\small
\begin{equation}
  \mathcal{I}_{64} = \left\{\begin{array}{@{}l@{\;}l@{}}
    \{u64 \cap s64_0, u64 \cap s64_1\} \setminus \{\emptyset\} & \text{if } s64 \text{ crosses } 0, \\
    \{u64 \cap s64\} \setminus \{\emptyset\} & \text{otherwise}
  \end{array}\right.
  \;
  \mathcal{I}_{32} = \left\{\begin{array}{@{}l@{\;}l@{}}
    \{u32 \cap s32_0, u32 \cap s32_1\} \setminus \{\emptyset\} & \text{if } s32 \text{ crosses } 0, \\
    \{u32 \cap s32\} \setminus \{\emptyset\} & \text{otherwise}
  \end{array}\right.
  \label{eq:stage1_64}
\end{equation}}%

If $\mathcal{I}_{64} = \emptyset$ or $\mathcal{I}_{32} = \emptyset$,
the signed and unsigned bounds are mutually contradictory, and the
procedure terminates immediately returning \texttt{false}.

\Para{Stage 2: Subspace projection and 32-bit bound intersection.}
As shown in Figure~\ref{fig:subspace_decomp}, each 64-bit
interval $I = [x, y] \in \mathcal{I}_{64}$ (with $x_{\mathrm{hi}} =
x_{[63:32]}$, $x_{\mathrm{lo}} = x_{[31:0]}$, $y_{\mathrm{hi}} =
y_{[63:32]}$, and $y_{\mathrm{lo}} = y_{[31:0]}$) decomposes across
32-bit subspace boundaries into at most three candidate pairs $(I_\pi,
K_\pi)$, where $I_\pi \subseteq \mathbb{W}_{32}$ is a projected 32-bit
interval and $K_\pi \subseteq \mathbb{W}_{32}$ is the interval of
upper 32-bit subspace prefixes:
\begin{enumerate}[leftmargin=1.5em,nosep]
  \item \emph{Left Boundary Candidate $(I_{\mathrm{left}},
  K_{\mathrm{left}})$:} If $x_{\mathrm{hi}} = y_{\mathrm{hi}}$,
    $I_{\mathrm{left}} = [x_{\mathrm{lo}}, y_{\mathrm{lo}}]$ with
    singleton prefix $K_{\mathrm{left}} = [x_{\mathrm{hi}},
      x_{\mathrm{hi}}]$ is the sole candidate; if $x_{\mathrm{hi}} <
    y_{\mathrm{hi}}$, $I_{\mathrm{left}} = [x_{\mathrm{lo}}, 2^{32} -
      1]$ with $K_{\mathrm{left}} = [x_{\mathrm{hi}},
      x_{\mathrm{hi}}]$.
  \item \emph{Intermediate Full Candidate $(I_{\mathrm{mid}},
  K_{\mathrm{mid}})$:} Constructed only when $y_{\mathrm{hi}} >
    x_{\mathrm{hi}} + 1$, with full 32-bit span $I_{\mathrm{mid}} =
    [0, 2^{32} - 1]$ and prefix range $K_{\mathrm{mid}} =
    [x_{\mathrm{hi}} + 1, y_{\mathrm{hi}} - 1]$.
  \item \emph{Right Boundary Candidate $(I_{\mathrm{right}},
  K_{\mathrm{right}})$:} Constructed only when $y_{\mathrm{hi}} >
    x_{\mathrm{hi}}$, with interval $I_{\mathrm{right}} = [0,
      y_{\mathrm{lo}}]$ and singleton prefix $K_{\mathrm{right}} =
    [y_{\mathrm{hi}}, y_{\mathrm{hi}}]$.
\end{enumerate}
Since $|\mathcal{I}_{64}| \le 2$, decomposing every $I \in
\mathcal{I}_{64}$ yields a set $\mathcal{I}_\pi$ of at most $2 \times
3 = 6$ projected candidates:

{
\begin{equation}
\mathcal{I}_\pi = \bigcup_{I \in \mathcal{I}_{64}} \big\{(I_{\mathrm{left}}, K_{\mathrm{left}}),\; (I_{\mathrm{mid}}, K_{\mathrm{mid}}),\; (I_{\mathrm{right}}, K_{\mathrm{right}})\big\} \label{eq:stage2_proj} \\
\end{equation}
}

We then intersect each projected 32-bit
slice $I_\pi$ with each target 32-bit interval $I_{32} \in
\mathcal{I}_{32}$ ($|\mathcal{I}_{32}| \le 2$) via standard interval
arithmetic, producing a narrowed candidate set $\mathcal{J}$ of at
most $|\mathcal{J}| \le 6 \times 2 = 12$ non-empty pairs:

{
\begin{equation}
  \mathcal{J} = \big\{(I_\pi \cap I_{32},\; K_\pi) \;\big|\; (I_\pi, K_\pi) \in \mathcal{I}_\pi,\; I_{32} \in \mathcal{I}_{32},\; I_\pi \cap I_{32} \neq \emptyset\big\} \label{eq:stage2_narrow}
\end{equation}
}

\Para{Stage 3: Tristate number stepping.}
Having reduced all four numerical interval domains into the candidate
set $\mathcal{J}$, the multi-domain query reduces to testing whether
any candidate $(I_\pi, K_\pi) \in \mathcal{J}$ intersects the tristate
number $t \in \mathbb{A}_{\text{tnum}}$, decomposed into upper 32 bits
$t_{\mathrm{hi}}$ and lower 32 bits $t_{\mathrm{lo}}$. For each
candidate $(I_\pi, K_\pi) \in \mathcal{J}$, we perform two checks:
\begin{enumerate}[leftmargin=1.5em,nosep]
  \item \textbf{Prefix compatibility}. The upper 32-bit prefix
    interval $K_\pi = [k_{\min}, k_{\max}]$ must intersect
    $t_{\mathrm{hi}}$. For boundary candidates $I_{\mathrm{left}}$ and
    $I_{\mathrm{right}}$ with singleton prefixes $K_\pi = [k, k]$,
    this requires a single membership test $(k \mathbin{\&}
    \mathord{\sim} t_{\mathrm{hi}}.m) = t_{\mathrm{hi}}.v$. For the
    intermediate candidate $I_{\mathrm{mid}}$, we test whether $K_\pi
    = [k_{\min}, k_{\max}]$ intersects $t_{\mathrm{hi}}$ via
    Theorem~\ref{thm:one_step} (one-step interval
    intersection). 

  If the prefix check fails, the candidate is discarded.
  \item \textbf{Lower 32-bit tristate intersection}. For each
    candidate passing the prefix check, we test whether its projected
    32-bit interval $I_\pi = [a, b]$ intersects $t_{\mathrm{lo}}$ by
    applying Theorem~\ref{thm:one_step} (one-step
    interval intersection). 
\end{enumerate}

If both checks succeed for any candidate in $\mathcal{J}$, a common
64-bit register value exists ($R \in CM$) and
\texttt{reg\_bounds\_intersect()} immediately returns \texttt{true};
if all candidates fail, it returns \texttt{false} ($R \in
EI$). Figure~\ref{fig:walkthrough} traces all three stages on two
example register states.

\begin{figure}[t]
\centering
\fbox{\begin{minipage}{0.96\textwidth}
\footnotesize
\textbf{Example 1: Non-empty intersection.}
Consider an input register state with $u64 = [1, 2 \cdot 2^{32} + 5]$, $s64 = [-10, 2 \cdot 2^{32} + 5]$, $u32 = [10, 20]$, $s32 = [10, 20]$, and tnum $t$ where upper 32 bits are odd and lower 32 bits are multiples of 16 ($t.v = \mathtt{0x0000000100000000}$, $t.m = \mathtt{0xfffffffefffffff0}$):
\begin{itemize}[leftmargin=1.4em,nosep]
  \item \textbf{Stage 1:} $s64$ crosses zero, splitting into $[0, 2 \cdot 2^{32} + 5] \cup [2^{64} - 10, 2^{64} - 1]$; intersecting with $u64$ gives $\mathcal{I}_{64} = \{[1, 2 \cdot 2^{32} + 5]\}$. $s32$ does not cross zero; intersecting with $u32$ gives $\mathcal{I}_{32} = \{[10, 20]\}$.
  \item \textbf{Stage 2:} $[1, 2 \cdot 2^{32} + 5]$ has $x_{\mathrm{hi}} = 0, x_{\mathrm{lo}} = 1, y_{\mathrm{hi}} = 2, y_{\mathrm{lo}} = 5$ ($y_{\mathrm{hi}} > x_{\mathrm{hi}} + 1$), decomposing into three slices intersected with $\mathcal{I}_{32}$:
  $\bullet$~Left: $[1, 2^{32} - 1] \cap [10, 20] \implies ([10, 20], [0, 0])$;
  $\bullet$~Middle: $[0, 2^{32} - 1] \cap [10, 20] \implies ([10, 20], [1, 1])$;
  $\bullet$~Right: $[0, 5] \cap [10, 20] = \emptyset$ (discarded).
  Surviving candidates: $\mathcal{J} = \{([10, 20], [0, 0]), ([10, 20], [1, 1])\}$.
  \item \textbf{Stage 3:}
  $\bullet$~Candidate 1 ($K_\pi = [0, 0]$): $0 \notin \gamma_{\text{tnum}}(t_{\mathrm{hi}})$ since $t_{\mathrm{hi}}$ requires odd integers (discarded).
  $\bullet$~Candidate 2 ($K_\pi = [1, 1]$): $1 \in \gamma_{\text{tnum}}(t_{\mathrm{hi}})$ passes the prefix check; for $I_\pi = [10, 20]$, $10 \notin \gamma_{\text{tnum}}(t_{\mathrm{lo}})$, but $\mathtt{tnum\_step}(t_{\mathrm{lo}}, 10) = 16 \le 20$. Candidate 2 intersects $t$, so the algorithm returns \textbf{\texttt{true}}.
\end{itemize}
\textbf{Example 2: Empty intersection.}
For the same state with $t_{\mathrm{lo}}$ requiring multiples of 32, $\mathtt{tnum\_step}(t_{\mathrm{lo}}, 10) = 32 > 20$; Candidate 2 fails, leaving no valid candidates, so the algorithm returns \textbf{\texttt{false}}.
\end{minipage}}
\vspace{-0.4em}
\caption{\textbf{End-to-End Walkthrough of Allwise Intersection.}}
\vspace{-0.4em}
\label{fig:walkthrough}
\end{figure}

\subsection{Implementation}
\label{sec:ebpf:simulation}
We implement our allwise intersection procedure in the Linux kernel's
conditional branch feasibility analysis to prune dead branches and
remove domain invariant violations before path exploration. 
On a conditional jump (\texttt{if (r1 op r2) goto target}) with an
undetermined outcome, the stock kernel verifier forks first and
refines \texttt{r1} and \texttt{r2} along each branch before invoking
\texttt{reg\_bounds\_sync()} (\S\ref{sec:background}). As shown in
\S\ref{sec:verifier}, refining operands on an infeasible branch
produces $\mathit{EI}$ states that may subsequently turn into
$\mathit{IF}$, triggering invariant violations and bound resets, or
states that may silently recover to well-formed states ($\mathit{EI}
\to \mathit{CM}$), causing path explosion and false rejections.

Our implementation prevents branches reaching $\mathit{EI} \cup
\mathit{IF}$ states from being explored in the first place. Before
forking, we check branch feasibility on refined scratch operand states:
(1) we clone the branch operands into stack-allocated scratch
registers ($R_{1t}, R_{1f} \leftarrow R_1$ and $R_{2t}, R_{2f}
\leftarrow R_2$) and refine them under the taken and fall-through
conditions; (2) we run our $O(1)$ allwise intersection check on each
refined scratch state; and (3) if any refined operand is in
$\mathit{EI} \cup \mathit{IF}$, Theorem~\ref{thm:dead_branch}
certifies the branch is unreachable and the verifier prunes it.

Pruning dead branches before forking ensures domain reduction and
downstream operators are never invoked on contradictory abstract
states ($R \in \mathit{EI}$). This not only eliminates domain
invariant violations (no more $\mathit{EI} \to \mathit{IF}$
transitions), but also prunes unreachable paths in the program,
allowing the kernel verifier more room to stay within its
budget of 1M checked instructions.

  \section{Experimental Evaluation}
\label{sec:eval}

We evaluate our allwise intersection check for pruning dead paths and
answer the following research questions. (Q1) Does the allwise
intersection check help detect historical invariant violations? (Q2)
Does the early dead-path pruning using allwise intersection checks
prevent state explosion? (Q3) Does it eliminate false rejections of
safe eBPF programs? (Q4) What is the verification time overhead and
performance cost on real-world production applications?

\Para{Methodology and setup.}  We conduct all our experiments on an
x86\_64 Linux host (24-core AMD EPYC 7352, 256\,GiB RAM), using a
16\,GiB QEMU x86\_64 guest for bug reproducers. We compare two kernel
configurations. (1)~\textbf{Stock baseline}, which is an unmodified
Linux kernel (\texttt{v7.1-rc1}, commit \texttt{7e85ca02ef3aa}) that
includes standard domain operators, pairwise reduction
(\texttt{reg\_bounds\_sync()}), and well-formedness checks
(\texttt{reg\_bounds\_sanity\_check()}) but lacks our intersection
check and pre-path exploration refinement and dead-path pruning.
(2)~\textbf{Our proposed system}, which is the same kernel augmented
with \texttt{tnum\_step()}, \texttt{reg\_bounds\_intersect()}, and
pre-path exploration branch refinement and pruning. On production
suites, we use \texttt{veristat} to record verified instructions
(\texttt{insns\_processed}), total states, and wall-clock verification
time ($\mu$s); micro-benchmarks and bug reproducers are loaded via a
custom C/\texttt{libbpf} harness (\texttt{log\_level=2}).

\Para{Ability to detect and prevent historical domain invariant
  violations (Q1).} We first evaluate whether our allwise intersection
check resolves the recurring class of domain invariant violations
affecting mainline Linux kernels. Over several years, kernel
developers addressed these violations reactively by patching
individual operator or reduction edge
cases. Table~\ref{tab:historical} lists five prominent upstream
bug-fix commits across \texttt{v6.17-rc1}--\texttt{v7.0-rc3}. We
loaded the minimal BPF bytecode reproducer for each commit into both
the stock kernel (with the point-fix reverted to reproduce the
pre-patch failure) and proposed kernels. As shown in
Table~\ref{tab:historical}, the stock verifier triggers \texttt{domain
  invariant violation} warnings across all five programs when
\texttt{reg\_bounds\_sync()} reduces contradictory domain bounds along
an infeasible branch.
In contrast, our proposed verifier invokes
\texttt{reg\_bounds\_intersect()} before exploring paths and detects
that no concrete 64-bit word satisfies all five domains. It
subsequently prunes the dead path immediately with zero warnings,
confirming that a single sound and complete intersection check
subsumes years of ad-hoc point fixes. In summary, our algorithm fixes
the root cause of these bugs!

\begin{table}[t]
\centering
\footnotesize
\setlength{\tabcolsep}{3pt}
\begin{tabular}{@{}lllllcc@{}}
\toprule
\textbf{Commit} & \textbf{Kernel} & \textbf{Upstream Patch Summary} & \textbf{Opcode} & \textbf{Conflict} & \textbf{Stock} & \textbf{Proposed} \\
\midrule
\texttt{fbc7aef5} & \texttt{7.0-rc3} & Fix u32/s32 bounds when ranges cross min/max & \texttt{JSLT} & $u32 \cap s32$ & \xmark & \cmark \\
\texttt{efc11a66} & \texttt{7.0-rc2} & Bounds when tnum has single values & \texttt{JEQ} & $\mathit{tnum} \cap u64$ & \xmark & \cmark \\
\texttt{f41345f4} & \texttt{6.18-rc1} & Use tnums for JEQ/JNE in \texttt{is\_branch\_taken} & \texttt{JNE} & $\mathit{tnum} \cap u64$ & \xmark & \cmark \\
\texttt{00bf8d0c} & \texttt{6.17-rc1} & Bounds when s64 crosses sign boundary & \texttt{JLT} & $s64 \cap u64$ & \xmark & \cmark \\
\texttt{6279846b} & \texttt{6.17-rc1} & Forget ranges when refining tnum after JSET & \texttt{JSET} & $u32 \cap u64$ & \xmark & \cmark \\
\bottomrule
\end{tabular}
\caption{\small Ability to detect historical invariant
  violations~(Q1). Our system detects the empty intersection and
  prunes the dead path (\cmark), whereas the stock verifier reports
  \texttt{domain invariant violation} (\xmark).}
\vspace{-1em}
\label{tab:historical}
\end{table}

\Para{Pruning infeasible paths avoids state space explosion (Q2).}
When the verifier fails to prune an infeasible path, it must explore
all downstream instructions along that infeasible path. The infeasible
path often contains bounded loops (\eg, packet parsing) or cascading
conditionals (\eg, firewall rule tables). If the cumulative
exploration exceeds the one million instruction limit
(\texttt{BPF\_COMPLEXITY\_LIMIT\_INSNS}), loading aborts with
the error code \texttt{-E2BIG}.

To evaluate the scalability benefits of early dead-branch detection,
we ground our evaluation in a known upstream reproducer, where the
\code{true} (taken) branch of a conditional jump at instruction~8
(\code{if w6 == 0x1 goto pc+1}) is infeasible due to an empty domain
intersection ($\mathit{tnum} \cap \mathit{u64} = \emptyset$). We
construct two parameterized microbenchmark suites
(Listings~\ref{lst:q2_workloads} and~\ref{lst:q2_trees}) attached to
the dead path:

\begin{minipage}[t]{0.42\textwidth}
\begin{lstlisting}[style=bpfstyle,basicstyle=\ttfamily\footnotesize,firstnumber=11,caption={Bounded loop ($K$ iterations).},label={lst:q2_workloads}]
r3 = K
r4 = 0
r4 += 1
r3 -= 1
if r3 s> 0x0 goto pc-3
\end{lstlisting}
\end{minipage}\hfill
\begin{minipage}[t]{0.54\textwidth}
\begin{lstlisting}[style=bpfstyle,basicstyle=\ttfamily\footnotesize,firstnumber=11,caption={Decision tree ($N$ diamond stages).},label={lst:q2_trees}]
r0 = bpf_get_prandom_u32(); // stage d in 0..N-1
if (r0 > 0x7fffffff) goto skip_d;
r7 += (1 << d);
skip_d: ... // repeated for N stages
if (r7 == 0x7fffffff) goto exit;
\end{lstlisting}
\end{minipage}

\Para{Bounded Loops ($K \in [10, 350{,}000]$).}  The bounded loop
benchmark (Listing~\ref{lst:q2_workloads}) simulates iterative packet
scanning along the dead path, initializing counter \texttt{r3 = K} and
accumulator \texttt{r4 = 0} and executing a 3-instruction loop body
(\texttt{add}, \texttt{sub}, \texttt{jsgt}) until \texttt{r3} reaches
$0$. As shown in Table~\ref{tab:scalability}(a), the Stock verifier
unrolls every iteration symbolically in $3K + 13$ instructions; at $K
= 350{,}000$ ($1{,}050{,}013$ instructions), it exceeds the 1M limit
and rejects the program with \texttt{-E2BIG}. The Proposed verifier
proves $\mathit{tnum} \cap u64 = \emptyset$ at the guarding jump,
pruning the dead loop and verifying in a constant $10$ instructions
across all $K$.

\Para{Branching Decision Trees ($N \in [1, 18]$).}  The decision
tree benchmark (Listing~\ref{lst:q2_trees}) models a classification
DAG of $N$ sequential diamond branches on the dead path. Each stage $d
\in [0, N - 1]$ branches on \texttt{bpf\_get\_prandom\_u32()} and
conditionally adds bit $2^d$ to accumulator \texttt{r7}, generating
$2^N$ distinct abstract values ($0 \le \mathtt{r7} < 2^N$) that
prevent state merging. As shown in Table~\ref{tab:scalability}(b),
Stock exploration grows exponentially ($O(2^N)$) from $20$
instructions ($N = 1$) to $393{,}224$ ($N = 16$), exhausting the 1M
limit at $N = 18$ ($262{,}144$ paths) with \texttt{-E2BIG}. The
Proposed verifier prunes the unreachable entry branch immediately,
verifying in $10$ instructions across all $N \in [1, 18]$.

\begin{table}[t]
\centering
\footnotesize
\setlength{\tabcolsep}{3.5pt}
\begin{minipage}[t]{0.48\textwidth}
\centering
\textbf{(a) Bounded Loops}\\[2pt]
\begin{tabular}{rrrrrc}
\toprule
\textbf{Bound} & \textbf{Scale} & \multicolumn{2}{c}{\textbf{Stock}} & \multicolumn{2}{c}{\textbf{Proposed}} \\
\cmidrule(lr){3-4} \cmidrule(lr){5-6}
$K$ & \textbf{(Iters)} & \textbf{Insns} & \textbf{Status} & \textbf{Insns} & \textbf{Status} \\
\midrule
$10$ & $10$ & $43$ & \cmark & $10$ & \cmark \\
$100$ & $100$ & $313$ & \cmark & $10$ & \cmark \\
$1{,}000$ & $1{,}000$ & $3{,}013$ & \cmark & $10$ & \cmark \\
$10{,}000$ & $10{,}000$ & $30{,}013$ & \cmark & $10$ & \cmark \\
$50{,}000$ & $50{,}000$ & $150{,}013$ & \cmark & $10$ & \cmark \\
$350{,}000$ & $350{,}000$ & $>1\text{M}$ & \textcolor{cred}{\texttt{-E2BIG}} & $10$ & \cmark \\
\bottomrule
\end{tabular}
\end{minipage}\hfill
\begin{minipage}[t]{0.50\textwidth}
\centering
\textbf{(b) Branching Decision Trees}\\[2pt]
\begin{tabular}{rrrrrc}
\toprule
\textbf{Depth} & \textbf{Scale} & \multicolumn{2}{c}{\textbf{Stock}} & \multicolumn{2}{c}{\textbf{Proposed}} \\
\cmidrule(lr){3-4} \cmidrule(lr){5-6}
$N$ & \textbf{(Paths)} & \textbf{Insns} & \textbf{Status} & \textbf{Insns} & \textbf{Status} \\
\midrule
$1$ & $2$ & $20$ & \cmark & $10$ & \cmark \\
$2$ & $4$ & $32$ & \cmark & $10$ & \cmark \\
$4$ & $16$ & $104$ & \cmark & $10$ & \cmark \\
$8$ & $256$ & $1{,}544$ & \cmark & $10$ & \cmark \\
$12$ & $4{,}096$ & $24{,}584$ & \cmark & $10$ & \cmark \\
$16$ & $65{,}536$ & $393{,}224$ & \cmark & $10$ & \cmark \\
$18$ & $262{,}144$ & $>1\text{M}$ & \textcolor{cred}{\texttt{-E2BIG}} & $10$ & \cmark \\
\bottomrule
\end{tabular}
\end{minipage}
\vspace{0.2em}
\caption{\small \textbf{Scalability and complexity limit avoidance
    under stock vs.\ our proposed verifiers (Q2)}. Bounded loops (a)
  scale linearly and decision trees (b) scale exponentially,
  exhausting the 1M instruction limit (\texttt{-E2BIG}) in the stock
  verifier, whereas our verifier prunes the dead path in 10
  instructions across all scales.}
\vspace{-0.5em}
\label{tab:scalability}
\end{table}

\Para{Elimination of false rejections and improvements in
  expressivity (Q3).}
We next evaluate whether our approach eliminates \emph{false
rejections} of safe eBPF programs containing unreachable code. In
Clang/LLVM-compiled programs, defensive error guards, inlined helpers,
and \texttt{switch} fallbacks often create branches impossible under
earlier constraints. When compiler optimizations obscure correlations
across checks, the Stock verifier explores the dead path, which may
leave a register uninitialized, compute an out-of-bounds offset, or
set a pointer to \texttt{NULL} before merging into shared code,
causing false rejections.

To evaluate this systematically, we construct five diamond-CFG
micro-benchmarks (P1--P5 in Table~\ref{tab:expressiveness}) spanning
five verifier safety domains, where the live path performs valid
initialization while the dead path ($\mathit{tnum} \cap u64 =
\emptyset$) violates a safety requirement before or at the merge
point. \textbf{(P1) Stack memory safety}: The live path assigns valid
stack offset \texttt{r8 = -8}; the dead path assigns out-of-bounds
\texttt{r8 = -600} before accessing \texttt{r10 + r8} at the merge
point. \textbf{(P2) Pointer safety}: The live path sets \texttt{r8 =
  r10}; the dead path sets \texttt{NULL} (\texttt{r8 = 0}) before
dereferencing \texttt{*(u64*)(r8 - 8)} at the merge point.
\textbf{(P3) Register initialization}: The live path initializes
\texttt{r8 = 42}; the dead path skips initialization and jumps to the
merge point reading \texttt{r0 = r8}. \textbf{(P4) Kernel context
  isolation}: The dead path reads offset 1000 from \texttt{skb}
(\texttt{*(u32*)(skb + 1000)}), exceeding \texttt{struct
  \_\_sk\_buff}. \textbf{(P5) Helper argument safety}: The live path
sets \texttt{r1 = skb}; the dead path sets scalar \texttt{r1 = 0}
before calling \texttt{bpf\_get\_socket\_cookie(r1)}.

As shown in Table~\ref{tab:expressiveness}, the stock verifier
explores the dead branch and falsely rejects all five programs,
whereas our proposed verifier detects $\mathit{tnum} \cap u64 =
\emptyset$ via \texttt{reg\_bounds\_intersect()}, prunes the dead
branch, and accepts all five safe programs.

\begin{table}[t]
\centering
\footnotesize
\setlength{\tabcolsep}{2.2pt}
\begin{tabular}{@{}llllc@{}}
\toprule
\textbf{ID} & \textbf{Safety Domain} & \textbf{Dead-Path Operation} & \textbf{Stock (Error)} & \textbf{Proposed} \\
\midrule
P1 & Stack Memory & OOB stack write: \texttt{*(fp - 600) = 42} & \xmark\ \texttt{invalid write to stack off=-600} & \cmark \\
P2 & Pointer Safety & Null dereference: \texttt{*(r8 - 8) = r0} & \xmark\ \texttt{R8 invalid mem access 'scalar'} & \cmark \\
P3 & Register Init. & Uninitialized read: \texttt{r0 = r8} & \xmark\ \texttt{R8 !read\_ok} & \cmark \\
P4 & Context Isol. & OOB context: \texttt{r2 = *(skb + 1000)} & \xmark\ \texttt{invalid bpf\_context access off=1000} & \cmark \\
P5 & Helper Args & Invalid arg: \texttt{bpf\_get\_socket\_cookie(0)} & \xmark\ \texttt{R1 type=scalar expected=ctx} & \cmark \\
\bottomrule
\end{tabular}
\caption{\small \textbf{Verifier program acceptance across five verifier
    safety domains (Q3).} Unsafe operations on an unreachable branch
  trigger false rejections in Stock (\xmark), whereas Proposed prunes
  the dead branch and accepts all five safe programs (\cmark).}
\vspace{-0.5em}
\label{tab:expressiveness}
\end{table}

\Para{Performance on real-world production applications (Q4).}  To
confirm that our approach maintains verifier throughput and does not
introduce regressions on complex real-world software, we evaluate our
system across two industrial benchmark suites. (1) \textbf{Linux
  kernel selftests}, comprising 3,334 program configurations across
742 compiled ELF files. (2) \textbf{Cilium complexity CI tests}
comprising 2,229 program configurations across 71 programs. Together,
this evaluation covers 5,563 test cases and over 40.3 million verified
instructions.

\begin{table}[b]
\centering \footnotesize
\begin{tabular}{lrrrrr}
\toprule
\textbf{Workload Suite} & \textbf{Configs} & \multicolumn{2}{c}{\textbf{Insns.\ Explored}} & \multicolumn{2}{c}{\textbf{Avg.\ Time ($\mu$s)}} \\
\cmidrule(lr){3-4} \cmidrule(lr){5-6}
 & & \textbf{Stock} & \textbf{Proposed} & \textbf{Stock} & \textbf{Proposed} \\
\midrule
Linux Kernel Selftests & $3{,}334$ & $8{,}718{,}982$ & $8{,}718{,}982$ ($+0.00\%$) & $20{,}784$ & $21{,}328$ ($+2.62\%$) \\
Cilium Complexity CI Tests & $2{,}229$ & $31{,}664{,}778$ & $31{,}666{,}493$ ($+0.01\%$) & $173{,}711$ & $175{,}928$ ($+1.28\%$) \\
\midrule
\textbf{Combined} & $\mathbf{5{,}563}$ & $\mathbf{40{,}383{,}760}$ & $\mathbf{40{,}385{,}475}$ ($\mathbf{+0.004\%}$) & $\mathbf{82{,}059}$ & $\mathbf{83{,}275}$ ($\mathbf{+1.48\%}$) \\
\bottomrule
\end{tabular}
\caption{\small \textbf{Verification performance across workloads
    (Q4)} over 5,563 BPF program configurations.}
\vspace{-0.5em}
\label{tab:production}
\end{table}

\Para{Soundness and regressions.}  Across all $5{,}563$
configurations, the proposed verifier with our $O(1)$ intersection
check achieves a $100\%$ verdict match against the stock kernel: in
Linux kernel selftests, all $1{,}820$ valid programs pass and all
$1{,}514$ negative tests are rejected; in Cilium, all $2{,}229$
configurations pass, confirming zero regressions or false rejections.

\Para{Control-flow and complexity impact.}  Across all $3{,}334$
Linux Kernel Selftests, instruction complexity is unchanged ($\Delta =
+0.00\%$). In Cilium, complexity is identical on $88.1\%$ of test
cases ($1{,}964$ configs). On the remainder, our verifier walks $14$
fewer instructions on \texttt{tail\_nodeport\_ipv4\_dsr} ($51$
configs) and \texttt{tail\_policy\_denied\_ipv4} ($18$ configs) by
pruning dead error-handling paths, and $+11$ to $+29$ instructions out
of $205{,}000$ ($\le 0.01\%$) on \texttt{cil\_to\_netdev}: as noted by
kernel developers~\cite{chaignon2025jeqjne}, pruning an upstream
branch can tighten the state cached at a merge point, requiring a few
extra instructions before state equivalence pruning matches. Across
all $31.66\text{M}$ Cilium instructions, the net complexity delta is
$+0.0054\%$.

\Para{Verification Throughput and Latency Overhead.}  Across more
than $40.38\text{M}$ verified instructions, the wall-clock
verification time overhead of executing
\texttt{reg\_bounds\_intersect()} is negligible: $+2.62\%$ on Linux
kernel selftests (median $+0.43\%$) and $+1.28\%$ on Cilium (median
$+1.19\%$), for an overall average overhead of $\mathbf{+1.48\%}$.

  \section{Related Work}
\label{sec:related}

\Para{Circular numbers and tnums in the eBPF verifier.}  Building on
wrapped
intervals~\cite{wrapped-intervals-2012,shung-hsi-wrapped-2023},
Zingerman recently upstreamed circular numbers (\code{cnum64},
\code{cnum32})~\cite{zingerman-cnums-2026} into the eBPF verifier,
replacing the four signed/unsigned interval domains with modular arcs
$(\text{\code{base}}, \text{\code{size}})$ that wrap across zero
without losing precision and add an explicit $\bot$ sentinel
(\code{CNUM\_EMPTY}). However, \code{cnum}--\code{tnum} reduction
still exchanges only outer bounds, missing circular arcs that fall
inside interior \code{tnum} gaps (\S\ref{sec:step}).
As observed on the BPF mailing list, our
framework directly decides joint emptiness for $(\mathtt{cnum64},
\mathtt{cnum32}, \mathtt{tnum})$: $\mathtt{cnum64}$ and
$\mathtt{cnum32}$ unify signed and unsigned bounds to directly yield
Stage~1's $\le 2$ unsigned 64-bit intervals ($\mathcal{I}_{64}$) and
$\le 2$ unsigned 32-bit intervals ($\mathcal{I}_{32}$) (or
\texttt{CNUM\_EMPTY}). Because $\mathtt{cnum32}$ still induces
$2^{32}$ periodic discontinuities in $\mathbb{W}_{64}$, Stages~2 and~3
of \S\ref{sec:ebpf} then project $\mathcal{I}_{64}$ into
$\mathbb{W}_{32}$ and invoke \texttt{tnum\_step}
(Theorem~\ref{thm:one_step}) to decide joint emptiness in $O(1)$ time.
Kernel maintainers have adopted \texttt{tnum\_step} in reduction to
detect singleton intersections~\cite{chaignon2026singleton} and
tighten interval bounds~\cite{alexei-trim-2026}, while
Vero~\cite{sun2026tnumstep,sun2027vero} has optimized tnum\_step to
synthesize a 4-line branchless strength reduction.

\Para{eBPF verification, proof-carrying frameworks, and fuzzers.}
Prior work in this area has formalized and verified the verifier's
tristate~\cite{tnums-cgo22} and range
analyses~\cite{agni-cav23,agni-sas24,bhat-2022}, compared operator
precision via differential synthesis~\cite{agni-sas25}, and
synthesized abstract transformers via CEGIS in
Amurth~\cite{amurth-2022,amurth2-reduced-2026} and
Vayu~\cite{vayu-ebpf25}. All of these frameworks assume satisfiable
inputs ($R \in CM$), whereas we target the orthogonal problem of
detecting and pruning unsatisfiable states ($EI \cup IF$) on
infeasible paths. Userspace verifiers such as
PREVAIL~\cite{prevail-pldi19} (using Zones~\cite{zone-mine-2001}) omit
bitmasks and thus avoid interval-bitmask conflicts at the cost of
bit-level precision, while proof-carrying frameworks
(BCF~\cite{sun2025prove}, PCB~\cite{cai2026pcb},
ExoBPF~\cite{nelson:exobpf:lpc:2021}) offload checks to external SMT
solvers. Unlike verifier
fuzzers~\cite{sun2024validating,sun2024finding,lyu2025veritas,hung2024brf,buzzer-2024-branching-exploit}
that expose invariant violations empirically, our work eliminates
their root cause by construction within the existing in-kernel
domains.

\Para{Reduced products and bitwise abstract domains.}
Astr{\'e}e~\cite{astree-2005,mine-bitfield} and
Frama-C~\cite{framac-2012} along with LLVM's KnownBits analysis
approximate $\rho^*$ via Granger's pairwise
reduction~\cite{granger-1992} and communicate only through outer
bounds~\cite{taneja2020testing}, missing $\bot$ whenever an interval
lies inside an interior bitmask gap (\S\ref{sec:background}). Our
abstract step framework (Theorems~\ref{thm:one_step}
and~\ref{thm:sufficient}) provides the first $O(1)$ emptiness check
between intervals and non-convex domains without enumerating
elements. Closest in spirit are Warren's bit-twiddling
algorithms~\cite[Ch.~2, 4]{warren2013hackers} (e.g., \code{minOR},
\code{maxOR}), which step to the next bitmask match via carry-rippling
only when the starting value already satisfies the mask ($z \in
\gamma(t)$); \code{tnum\_step} generalizes this to any arbitrary
starting integer $z \in \mathbb{W}_{64}$ in $O(1)$ time.

  \section{Conclusion}
\label{sec:conclusion}

We expose why infeasible branches trigger domain invariant violations
in the Linux eBPF verifier: refining registers along dead paths
creates empty-intersection ($\bot$) states across interval and
non-convex bitwise domains that violate the implicit non-emptiness
precondition of domain reduction. We propose abstract step functions
and \code{tnum\_step}, a verified $O(1)$ bitwise stepping operator now
merged into the mainline Linux kernel. Using \code{tnum\_step} to
check multi-domain emptiness before branch exploration eliminates
dead-branch invariant violations at their root and resolves false
rejections in production workloads.

  \section{AI Use}
We used Gemini 4 Argon to develop the Lean formalization of
\texttt{tnum\_step} and allwise intersection check proofs and to
generate candidate test cases with a specified bound used for the
evaluation. The authors claim full responsibility for the ideas,
experimental design, research methodology, findings, and exposition in
this paper.

\begin{acks}
  We thank Yalu Cai for feedback on this draft. We also thank the
  Linux kernel developers and Paul Chaignon for feedback on various
  versions of the patches that we submitted.  This material is based
  upon work supported in part by the National Science Foundation
  FMITF-Track-II grant-2422076, the eBPF Foundation's research gift,
  and the GEMINI research award from Google Inc.
\end{acks}

\bibliographystyle{ACM-Reference-Format}
\bibliography{refs}

\end{document}